\documentclass[a4paper,fleqn]{cas-dc}

\usepackage[authoryear]{natbib}
\def\tsc#1{\csdef{#1}{\textsc{\lowercase{#1}}\xspace}}
\tsc{WGM}
\tsc{QE}
\begin{document}
\let\WriteBookmarks\relax
\def\floatpagepagefraction{1}
\def\textpagefraction{.001}

\shorttitle{TransLK-Net for Volumetric Medical Image Segmentation}    

\shortauthors{Jin Yang, Daniel S. Marcus and Aristeidis Sotiras}  

\title [mode = title]{TransLK-Net: Entangling Transformer and Large Kernel for Progressive and Collaborative Feature Encoding and Decoding in Medical Image Segmentation}  




\author[1]{Jin Yang}
\ead{yang.jin@wustl.edu}

\author[1]{Daniel S. Marcus}

\author[1,2]{Aristeidis Sotiras}


\ead{aristeidis.sotiras@wustl.edu}
\cortext[2]{Corresponding author}
\cormark[2]


\affiliation[1]{organization={Mallinckrodt Institute of Radiology},
            addressline={Washington University School of Medicine in St. Louis}, 
            city={St. Louis},
            postcode={63110}, 
            state={MO},
            country={USA}}

\affiliation[2]{organization={Institute for Informatics, Data Science and Biostatistics},
            addressline={Washington University School of Medicine in St. Louis}, 
            city={St. Louis},
            postcode={63110}, 
            state={MO},
            country={USA}}


\begin{abstract}
Convolutional neural networks (CNNs) and vision transformers (ViTs) are widely employed for medical image segmentation, but they are still challenged by their intrinsic characteristics. CNNs are limited from capturing varying-scaled features and global contextual information due to the employment of fixed-sized kernels. In contrast, ViTs employ self-attention and MLP for global information modeling, but they lack mechanisms to learn spatial-wise local information. Additionally, self-attention leads the network to show high computational complexity. To tackle these limitations, we propose Progressively Entangled Transformer Large Kernel (PTLK) and Collaboratively Entangled Transformer Large Kernel (CTLK) modules to leverage the benefits of self-attention and large kernel convolutions and overcome shortcomings. Specifically, PTLK and CTLK modules employ the Multi-head Large Kernel to capture multi-scale local features and the Efficient Decomposed Self-attention to model global information efficiently. Subsequently, they employ the Attention Entanglement mechanism to enable local and global features to enhance and calibrate each other progressively and collaboratively. Additionally, an Attention-gated Channel MLP (AG-MLP) module is proposed to equip the standard MLP module with the capabilities of modeling spatial information. PTLK and CTLK modules are further incorporated as a Cross Entanglement Decoding (CED) block for efficient feature fusion and decoding. Finally, we propose a novel network for volumetric medical image segmentation that employs an encoder-decoder architecture, termed TransLK-Net. The encoder employs a hierarchical ViT architecture whose block is built by incorporating PTLK and CTLK with AG-MLP into a ViT block, and the decoder employs the CED block. TransLK-Net was evaluated on four heterogeneous volumetric medical image segmentation tasks and achieved superior performance than other state-of-the-art methods with lower computational complexity. The code is made available at https://github.com/sotiraslab/TransLK-Net.

\end{abstract}


\begin{highlights}
    \item We propose novel Progressively Entangled Transformer Large Kernel and Collaboratively Entangled Transformer Large Kernel modules to leverage the benefits of large kernel convolutions and transformer self-attention for generic local and global feature extraction.
    \item We propose a novel Attention-gated Channel MLP to equip MLP with the capabilities of capturing pixel-wise spatial information, thus improving its representation powers for segmentation.
    \item We propose a Cross Entanglement Decoding block for efficient feature fusion and information decoding. It employs Progressive Feature Interaction and Collaborative Feature Interaction modules to fuse features from the decoder and skip-connected from the encoder, avoiding semantic gaps during fusion.
    \item We propose a novel network for volumetric medical image segmentation, termed TransLK-Net by incorporating these modules into a hierarchical ViT-based encoder-decoder architecture. It achieved superior performance with low computational complexity compared to other SOTA methods in four heterogeneous volumetric segmentation tasks.
\end{highlights}

%\nocite{*}

% Keywords

\maketitle
\section{Introduction}
Segmentation of organs and lesions in medical images supports clinical workflows, including diagnosis, prognosis, and treatment. However, manual annotation or contouring is labor-intensive and error-prone, motivating the development of automatic tools to improve efficiency and avoid errors \citep{wang2021annotation}. Deep learning (DL) techniques have revolutionized this field and various DL-based automatic segmentation methods have been developed with great success. Among these methods, convolutional neural networks (CNNs), especially U-Net \citep{ronneberger2015u} and its variants \citep{zhou2018unet,schlemper2019attention,huang2020unet,isensee2021nnu,yang2025dynamic}, are widely applied for automatic medical image segmentation. These methods are proposed for organ or tissue segmentation in various medical image modalities, e.g., Computed Tomography (CT) \citep{minnema2018ct,yang2023abdominal,zhou2024sbc,yang2025active}, Magnetic Resonance (MR) \citep{moeskops2016automatic,fu2018novel}, Positron Emission Tomography (PET) \citep{zhao2018tumor}, X-ray \citep{bullock2019xnet,konya2021convolutional}, and Electronic Microscopy (EM) \citep{oztel2017mitochondria,khadangi2021stellar}.

However, CNN-based methods are still challenged in various medical image segmentation tasks. First, CNNs are challenged to capture multi-scale features due to their employment of convolutional layers with a fixed-sized kernel \citep{cai2016unified}. Thus, they are limited from modeling complex structures and large inter-subject variations in shape and size, resulting in inferior segmentation performance. Several methods have been proposed to improve CNNs by equipping them with capabilities to capture multi-scale features, such as feature pyramid networks \citep{lin2017feature,seferbekov2018feature,zhao2019m2det} and multi-branch segmentation networks \citep{aslani2019multi,chen2022mbanet}. However, these prior works change the macro-architecture of CNNs and may not be readily integrated into general CNN structures, limiting their applicability. Other methods employ several convolutional layers with different kernel sizes to extract multi-scale features in the same level \citep{yin2023amsunet,alam2022multi,yang2025dmc}. However, features from the same channel are extracted repeatedly by multiple convolutional layers, leading to redundant use of computational resources and model parameters \citep{lin2023scale}. Second, CNNs fail to utilize global contextual information outside the receptive field to model long-range dependencies among features when convolutional layers extract features within local kernel windows \citep{bello2019attention}.

Compared with CNNs, Vision Transformers (ViTs) model long-range dependencies by employing the attention mechanism (e.g., self-attention and cross-attention) \citep{dosovitskiy2020image}. This mechanism empowers ViT-based models with image-level receptive fields to utilize global contextual information across the entire input image \citep{fan2024rmt}. However, ViT-based models are challenged by intrinsic characteristics of the attention mechanism to achieve superior segmentation performance in medical images. Specifically, self-attention processes 1D input sequences without localization information, lowering the capabilities of ViT-based models in capturing local features \citep{chen2024transunet}. Additionally, the attention mechanism has a high computational cost in high-resolution images, and medical images are predominantly 3D volumetric data with high resolutions, leading to a high computational complexity of volumetric medical image segmentation models \citep{lin2023scale}.

To tackle these limitations, we propose a \textbf{Progressively Entangled Transformer Large Kernel} (PTLK) module and a \textbf{Collaboratively Entangled Transformer Large Kernel} (CTLK) module to leverage the benefits of large kernel-based convolutional layers and self-attention from ViT and overcome their shortcomings. Specifically, PTLK and CTLK modules employ a Multi-head Large Kernel (MHLK) to capture multi-scale features with low computational complexity effectively. They utilize multiple convolutional layers with different large kernels to extract varying-scaled local features. Additionally, these convolutional layers are applied to different channels to avoid redundant feature extraction. Subsequently, PTLK and CTLK modules employ a Decomposed Efficient Self-attention (DESA) to adaptively model global information for channel features. This DESA is a more efficient way to calculate self-attention scores with lower computational complexity than standard self-attention. Finally, PTLK and CTLK employ an Attention Entanglement mechanism to aggregate features from MHLK and DESA, motivating the mutual enhancement between them. To improve the diversity of features, PTLK and CTLK implement this Attention Entanglement mechanism in two different ways. PTLK utilizes local features from MHLK to enhance global contextual information from DESA progressively. In contrast, CTLK utilizes local features from MHLK and global features from DESA to calibrate and enhance each other mutually. 

ViTs employ the feed-forward network (FFN) or multi-layer perceptron (MLP) module to introduce non-linearity to channels \citep{dosovitskiy2020image,xu2021co,liu2021swin}. However, these FFN and MLP modules limit the representation abilities of ViTs in medical image segmentation. Specifically, they lack mechanisms to capture pixel-wise features and model spatial contextual information due to the employment of fully connected layers, thus lowering segmentation accuracy \citep{chen2023dual}. To tackle this limitation, we propose a novel \textbf{Attention-gated Channel MLP} (AG-MLP) module to enhance its capabilities of capturing spatial information. AG-MLP employs a depth-wise convolutional layer to capture spatial-wise features efficiently, and a spatial-wise attention gate to calibrate these features with global spatial information, thus improving the representation powers of the MLP module.

The U-shaped encoder-decoder architecture is widely employed in medical image segmentation networks, and it adapts either concatenation or summation to fuse upsampled features in the decoder with those skip-connected from the encoder \citep{cao2022swin,lin2022ds,azad2024beyond}. However, this design ignores the semantic gap of features between the decoder and the encoder, thus lowering segmentation accuracy \citep{wang2022uctransnet}. To tackle this limitation, we propose a \textbf{Cross Entanglement Decoding} (CED) block for efficient feature fusion and decoding. It employs two parallel paths. The first path employs a \textbf{Progressive Feature Interaction} (PFI) module for progressive feature interaction and fusion, and subsequently, a CTLK block is utilized to decode finer features. In the second path, a \textbf{Collaborative Feature Interaction} (CFI) module is employed for collaborative feature interaction and fusion, and a PTLK block is followed to decode fused features.

We propose a novel network for volumetric medical image segmentation, termed \textbf{TransLK-Net}, by incorporating these modules into a hierarchical ViT-based encoder-decoder architecture. Specifically, to leverage the scaling behavior of hierarchical ViTs, we incorporate PTLK and CTLK modules along with AG-MLP into a standard ViT block, generating PTLK and CTLK blocks, respectively. The encoder employs the PTLK and CTLK blocks for generic feature extraction, and the decoder employs the CED block for feature fusion and decoding. To demonstrate the generalizability of TransLK-Net, we evaluated it on four heterogeneous volumetric segmentation tasks: abdominal multi-organ CT segmentation, brain tumor MR segmentation, hepatic vessel tumor CT segmentation, and abdomen organ CT segmentation. It achieved superior segmentation performance with lower computational complexity compared with other state-of-the-art (SOTA) models. Our contributions are summarized as follows:
\begin{itemize}
    \item We propose novel \textbf{Progressively Entangled Transformer Large Kernel} and \textbf{Collaboratively Entangled Transformer Large Kernel} modules to leverage the benefits of large kernel convolutions and transformer self-attention for generic local and global feature extraction.
    \item We propose a novel \textbf{Attention-gated Channel MLP} to equip MLP with the capabilities of capturing pixel-wise spatial information, thus improving its representation powers for segmentation.
    \item We propose a \textbf{Cross Entanglement Decoding} block for efficient feature fusion and information decoding. It employs Progressive Feature Interaction and Collaborative Feature Interaction modules to fuse features from the decoder and skip-connected from the encoder, avoiding semantic gaps during fusion.
    \item We propose a novel network for volumetric medical image segmentation, termed \textbf{TransLK-Net} by incorporating these modules into a hierarchical ViT-based encoder-decoder architecture. It achieved superior performance with low computational complexity compared to other SOTA methods in four heterogeneous volumetric segmentation tasks.
\end{itemize}

\section{Related work}
\subsection{Transformers in medical image segmentation}
The introduction of vision transformer has revolutionized the field of medical image segmentation \citep{dosovitskiy2020image}. Various ViT-based segmentation networks have been proposed with great success. Swin UNet was the first pure ViT-based network for medical image segmentation \citep{cao2022swin}. Medical Transformer utilized an axial-gated attention mechanism in Transformer blocks for local and global feature representation \citep{valanarasu2021medical}. DS-TransUNet employed duel Swin Transformer branches to capture features from patches at different scales \citep{lin2022ds}. MISSFormer was built by incorporating Enhanced Transformer blocks and the Enhanced Transformer Context Bridge into a hierarchical encoder-decoder network \citep{huang2022missformer}. nnFormer was proposed as 3D Transformer network for volumetric medical image segmentation \citep{zhou2023nnformer}. CSWin-UNet incorporated the cross-shaped window self-attention-based Transformer module into the UNet for efficient long-range dependencies modeling \citep{liu2024cswin}. AgileFormer utilized spatially dynamic components to effectively capture heterogeneous features from target organs \citep{qiu2026agileformer}.

\subsection{Hybrid CNN-Transformer networks in medical image segmentation}
To enhance the capabilities of segmentation models by leveraging the advantages of both CNNs and ViTs, some methods attempt to combine them by utilizing a ViT-based encoder and a CNN-based decoder. TransUNet utilized a Vision Transformer to encode tokenized feature maps for capturing global information and a CNN decoder to generate pixel-wise segmentation results \citep{chen2024transunet}. CoTr followed a similar design as TransUNet where a ViT encoder was employed to enhance features from the CNN encoder \citep{xie2021cotr}. UNETR and Swin UNETR employed a ViT and Swin ViT encoder respectively to learn hierarchical feature representations, and then these encoders were incorporated with a CNN decoder for segmentation \citep{hatamizadeh2022unetr,hatamizadeh2021swin}. TransFuse employed two parallel CNN and Transformer branches for information processing \citep{zhang2021transfuse}. CTC-Net employed a ViT and CNN encoder to capture complementary features for a CNN decoder \citep{yuan2023effective}. Similarly, CT-Net fused features from an asymmetric ViT encoder and a CNN encoder for global and local feature representations \citep{zhang2024ct}. Large kernel convolutional layers were incorporated into ViT architectures to utilize local and global contextual information \citep{yang2026d}. Dynamic MLP Mixer network employed attention-enhanced MLP modules in ViT architecture \citep{yang2025d2}

\subsection{Channel-wise and spatial-wise attention mechanisms}
Some attention mechanisms are proposed to enhance CNNs' capabilities of capturing long-range dependencies. The Squeeze and Excitation network was proposed to model inter-dependencies among global channels \citep{hu2018squeeze}. Subsequently, an Efficient Channel Attention mechanism was designed to improve channel attention learning by employing a local cross-channel interaction strategy \citep{wang2020eca}. The Squeeze and Attention network modeled spatial-wise attention by implementing convolution operations across pixels \citep{zhong2020squeeze}. The Convolutional Block Attention Module (CBAM) captured inter-channel and inter-spatial dependencies of features sequentially, thus enhancing feature representation learning \citep{woo2018cbam}. Channel-wise and spatial-wise attention were combined in parallel to capture global information among channels and spatial positions \citep{roy2018recalibrating}.

\section{Methods}
\subsection{Progressively Entangled Transformer Large Kernel}
\begin{figure*}[!t]
\centering
\includegraphics[width=0.9\textwidth]{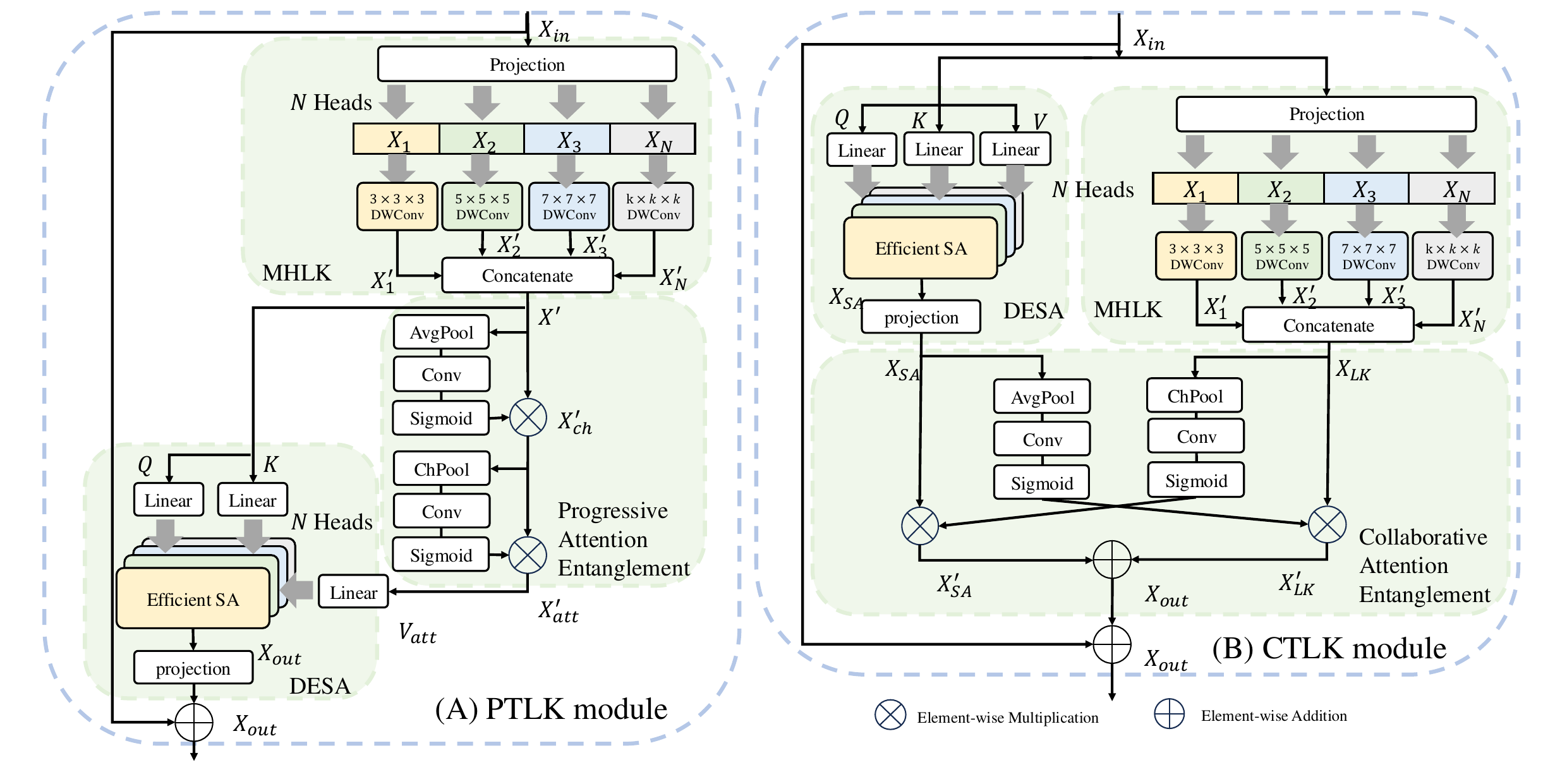}
\caption{The architecture of the (A) PTLK module and the (B) CTLK module. (A) In PTLK, MHLK projects the input features $X_{in}$ into $N$ heads as sub-features $\{X_1, X_2,..., X_N\}$ along channels. $N$ depth-wise large kernel convolutional layers (DWConv) are utilized to capture features $\{X_1', X_2',..., X_N'\}$. These features are concatenated as features $X'$, and Progressive Attention Entanglement is employed to enhance $X'$, generating features $X'_{att}$. DESA employs the linear projection to generate $Q$ and $K$ from $X'$ and $V$ from $X'_{att}$. After they are split into $N$ heads, the efficient self-attention (SA) mechanism is employed to calculate their attention scores, generating the output features $X_{out}$. Lastly, a residual connection is applied to $X_{in}$ and $X_{out}$. (B) In CTLK, MHLK projects the input features $X_{in}$ into $N$ heads as sub-features $\{X_1, X_2,..., X_N\}$ along channels. $N$ depth-wise large kernel convolutional layers are utilized to capture features $\{X_1', X_2',..., X_N'\}$. These features are concatenated as features $X_{LK}$. DESA employs the linear projection to create $Q$, $K$, and $V$ with $N$ heads from features $X_{in}$. The efficient self-attention mechanism is employed to calculate their attention scores, generating the output features $X_{SA}$. Subsequently, the Collaborative Attention Entanglement is employed to enhance features $X_{LK}$ and $X_{SA}$, generating features $X'_{LK}$ and $X'_{SA}$. Lastly, they are added to generate output features $X_{out}$, and a residual connection is applied to $X_{in}$ and $X_{out}$} 
\label{fig1}
\end{figure*}

The PTLK module employs three components: Multi-head Large Kernel (MHLK), Progressive Attention Entanglement, and Decomposed Efficient Self-attention (DESA) (Figure \ref{fig1}A).

\textbf{Multi-head Large Kernel.} MHLK partitions input features $\boldsymbol{X}_{in}\in\mathbb{R}^{C\times D\times H\times W}$ into $N$ heads as a set of sub-features $\{\boldsymbol{X}_1,\boldsymbol{X}_2,...,\boldsymbol{X}_N\}=\{\boldsymbol{X}_i\in\mathbb{R}^{\frac{C}{N}\times D\times H\times W}|i\in\{1,2,..., N\}\}$ along channels via a linear projection, where $C$ represents the number of channels, and $D$, $H$, and $W$ represent the depth, height, and width of volumetric images, respectively
\begin{align}
    \{\boldsymbol{X}_1,\boldsymbol{X}_2,...,\boldsymbol{X}_N\}&=\textrm{Projection}(\boldsymbol{X}_{in}).
\end{align}
Subsequently, $N$ distinct depth-wise large kernel convolutional layers ($\textrm{DWConv}_{k_i}$) are utilized to extract features in each head separately. These convolutional layers have growing kernel sizes $k_i\in\{k_1,k_2,...,k_N\}$, thus capturing varying-scaled features. We start from a convolution layer with the $3\times3\times3$ kernel size ($k_1=3$) and utilize convolutional layers whose kernel sizes gradually increase by 2 per head ($k_i=3+2*(i-1)$). Features from each head are concatenated to generate the output features $\boldsymbol{X}'\in\mathbb{R}^{C\times D\times H\times W}$:
\begin{equation} 
\begin{aligned}
    \boldsymbol{X}'_1&=\textrm{DWConv}_{k_1}(\boldsymbol{X}_1),\\
    \boldsymbol{X}'_2&=\textrm{DWConv}_{k_2}(\boldsymbol{X}_2),\\
    ...\\
    \boldsymbol{X}'_N&=\textrm{DWConv}_{k_N}(\boldsymbol{X}_N).
\end{aligned}
\end{equation}
Features $\boldsymbol{X}'_i\in\mathbb{R}^{\frac{C}{N}\times D\times H\times W}$ from each head are concatenated to generate the output features $\boldsymbol{X}'\in\mathbb{R}^{C\times D\times H\times W}$:
\begin{equation}
    \boldsymbol{X}'=\textrm{Concat}([\boldsymbol{X}'_1,\boldsymbol{X}'_2,...,\boldsymbol{X}'_N]).
\end{equation}
This design enables us to regulate the range of receptive fields and multi-granularity information without adjusting the number of heads. Additionally, these growing kernel sizes provide large receptive fields, thus capturing more long-range contextual information. Since different convolutional kernels are applied to different channels, rather than applying all kernels to each channel, redundant features from the same channel will not be captured.

\textbf{Progressive Attention Entanglement.} To entangle local features from MHLK to the following DESA, we utilize the Progressive Attention Entanglement mechanism to calibrate these local features $\boldsymbol{X}'$. First, channel-wise attention is employed by cascading an adaptive average pooling layer (AvgPool), a linear layer, and a Sigmoid function. It models the importance of these kernels and inter-dependencies among these local features $\boldsymbol{X}'$, generating features $\boldsymbol{X}'_{ch}\in\mathbb{R}^{C\times D\times H\times W}$
\begin{align}
    \boldsymbol{X}'_{ch}&=\boldsymbol{X}'*\textrm{Sigmoid}(\textrm{Linear}(\textrm{AvgPool}(\boldsymbol{X}'))).    
\end{align}
Subsequently, since DESA lacks mechanisms to learn spatial-wise representations, a spatial-wise attention is employed to enhance features $\boldsymbol{X}'_{ch}$ via spatial-wise global information, generating global information enhanced features $\boldsymbol{X}'_{att}\in\mathbb{R}^{C\times D\times H\times W}$ for DESA. This spatial-wise attention mechanism cascades a channel pooling layer (ChPool), a $7\times7\times7$ convolutional layer (Conv7), and a Sigmoid function
\begin{align}
    \boldsymbol{X}'_{att}&=\boldsymbol{X}'_{ch}*\textrm{Sigmoid}(\textrm{Conv7}(\textrm{ChPool}(\boldsymbol{X}'_{ch}))) .   
\end{align}

\textbf{Decomposed Efficient Self-attention (DESA).} DESA takes attention-entangled features $\boldsymbol{X}'_{att}$ as input to model their inter-dependencies and feature representations. A linear projection is applied to features $\boldsymbol{X}'_{att}$ to generate the value matrix $\boldsymbol{V}_{att}\in\mathbb{R}^{C\times D\times H\times W}$. The same linear projection is applied to the output features $\boldsymbol{X}'$ from MHLK to generate the query matrix $\boldsymbol{Q}\in\mathbb{R}^{C\times D\times H\times W}$ and the key matrix $\boldsymbol{K}\in\mathbb{R}^{C\times D\times H\times W}$
\begin{align}
    \{\boldsymbol{Q}, \boldsymbol{K}, \boldsymbol{V}_{att}\} = \textrm{Projection}(\{\boldsymbol{X}',\boldsymbol{X}',\boldsymbol{X}'_{att}\}).
\end{align}
Subsequently, these projected vectors are split into $N$ heads
\begin{equation} 
\begin{aligned}
    \boldsymbol{Q}&=\{Q_i\in\mathbb{R}^{\frac{C}{N}\times D\times H\times W}|i\in\{1,2,...,N\}\}\\
    &=\{Q_1,Q_2,...,Q_N\}\in\mathbb{R}^{N\times\frac{C}{N}\times D\times H\times W},\\
    \boldsymbol{K}&=\{K_i\in\mathbb{R}^{\frac{C}{N}\times D\times H\times W}|i\in\{1,2,...,N\}\}\\
    &=\{K_1,K_2,...,K_N\}\in\mathbb{R}^{N\times\frac{C}{N}\times D\times H\times W},\\
    \boldsymbol{V}_{att}&=\{(V_{att})_i\in\mathbb{R}^{\frac{C}{N}\times D\times H\times W}|i\in\{1,2,...,N\}\}\\
    &=\{(V_{att})_1,(V_{att})_2,...,(V_{att})_N\}\in\mathbb{R}^{N\times\frac{C}{N}\times D\times H\times W}.
\end{aligned}
\end{equation}
The self-attention score for $i$-th head is calculated as
\begin{align}
   (\boldsymbol{X}_{out})_i=\textrm{Softmax}\bigg(\frac{Q_i{K}_i^T}{\sqrt{d_m}}\bigg)(V_{att})_i.
\end{align}
where $d_m=\frac{C}{N}$. However, applying this self-attention mechanism to 4D input matrices leads to high computational complexity. To avoid high complexity, we introduce an efficient way to calculate self-attention scores by decomposing 4D matrices into lower dimensions. After decomposition, it calculates attention scores between $\boldsymbol{Q}$ and $\boldsymbol{K}$ along three dimensions, $H$, $W$, and $D$, separately
\begin{equation} 
\begin{aligned}
   (\boldsymbol{X}_H)_i=\textrm{Softmax}\bigg(\frac{(Q_H)_i({K}_H)_i^T}{\sqrt{d_m}}\bigg)(V_{att})_i, \\
   (\boldsymbol{X}_W)_i=\textrm{Softmax}\bigg(\frac{(Q_W)_i({K}_W)_i^T}{\sqrt{d_m}}\bigg)(V_{att})_i, \\
   (\boldsymbol{X}_{out})_i=\textrm{Softmax}\bigg(\frac{(Q_D)_i({K}_D)_i^T}{\sqrt{d_m}}\bigg)(V_{att})_i.
\end{aligned}
\end{equation}
After features from all heads are concatenated and reshaped, the output features $\boldsymbol{X}_{out}\in\mathbb{R}^{C\times D\times H\times W}$ are obtained via the linear projection and the drop-out operation. Finally, a residual connection is applied
\begin{equation} 
\begin{aligned}
    \boldsymbol{X}_{out}&=\textrm{Concat}([(\boldsymbol{X}_{out})_1, (\boldsymbol{X}_{out})_2,...,(\boldsymbol{X}_{out})_N]),\\
    \boldsymbol{X}_{out}&=\textrm{Dropout}(\textrm{Projection}(\boldsymbol{X}_{out})).
\end{aligned}
\end{equation}
The final stage of PTLK applies a residual connection
\begin{equation} 
    \boldsymbol{X}_{out}=\boldsymbol{X}_{out}+\boldsymbol{X}_{in}.
\end{equation}

\subsection{Collaboratively Entangled Transformer Large Kernel}
The CTLK module employs three components: Multi-head Large Kernel, Decomposed Efficient Self-attention, and Collaborative Attention Entanglement (Figure \ref{fig1}B).

\textbf{Multi-head Large Kernel.} MHLK in CTLK shares the same design as that in PTLK. The output features $\boldsymbol{X}_{LK}\in\mathbb{R}^{C\times D\times H\times W}$ are captured from input features $\boldsymbol{X}_{in}$ via the multiple convolutional layers with growing large kernel sizes along different channels
\begin{align}
    \boldsymbol{X}_{LK}=\textrm{MHLK}(\boldsymbol{X}_{in}).
\end{align}

\textbf{Decomposed Efficient Self-attention.} DESA in CTLK has the same design as that in PTLK. The linear projection is applied to the input features $\boldsymbol{X}_{in}$ to generate the query matrix $\boldsymbol{Q}\in\mathbb{R}^{C\times D\times H\times W}$, the key matrix $\boldsymbol{K}\in\mathbb{R}^{C\times D\times H\times W}$, and the value matrix $\boldsymbol{V}\in\mathbb{R}^{C\times D\times H\times W}$,
\begin{align}
    \{\boldsymbol{Q}, \boldsymbol{K}, \boldsymbol{V}\} = \textrm{Projection}(\{\boldsymbol{X}_{in},\boldsymbol{X}_{in},\boldsymbol{X}_{in}\}).
\end{align}
Subsequently, these projected vectors are split into $N$ heads
\begin{equation} 
\begin{aligned}
    \boldsymbol{Q}&=\{Q_1,Q_2,...,Q_N\},\\
    \boldsymbol{K}&=\{K_1,K_2,...,K_N\},\\
    \boldsymbol{V}&=\{V_1,V_2,...,V_N\}.
\end{aligned}
\end{equation}
The self-attention scores are calculated efficiently in the same way 
\begin{equation} 
\begin{aligned}
   (\boldsymbol{X}_W)_i=\textrm{Softmax}\bigg(\frac{(Q_W)_i({K}_W)_i^T}{\sqrt{d_m}}\bigg)V_i, \\
   (\boldsymbol{X}_H)_i=\textrm{Softmax}\bigg(\frac{(Q_H)_i({K}_H)_i^T}{\sqrt{d_m}}\bigg)V_i, \\
   (\boldsymbol{X}_{SA})_i=\textrm{Softmax}\bigg(\frac{(Q_D)_i({K}_D)_i^T}{\sqrt{d_m}}\bigg)V_i.
\end{aligned}
\end{equation}
After features from all heads are concatenated, the output features $\boldsymbol{X}_{SA}\in\mathbb{R}^{C\times D\times H\times W}$ are obtained via the linear projection and the drop-out operation
\begin{equation} 
\begin{aligned}
    \boldsymbol{X}_{SA}&=\textrm{Concat}([(\boldsymbol{X}_{SA})_1, (\boldsymbol{X}_{SA})_2,...,(\boldsymbol{X}_{SA})_N]),\\
    \boldsymbol{X}_{SA}&=\textrm{Dropout}(\textrm{Projection}(\boldsymbol{X}_{SA})).
\end{aligned}
\end{equation}

\textbf{Collaborative Attention Entanglement.} Collaborative Attention Entanglement employs channel-wise and spatial-wise attention to improve mutual calibration between local features from MHLK and global features from DESA, thus improving representation capabilities collaboratively. Specifically, MHLK captures varying-scaled spatial local features, while DESA lacks mechanisms to capture spatial-wise localization information. Thus, spatial-wise attention maps are extracted from local features of MHLK $\boldsymbol{X}_{LK}$ to demonstrate the important spatial contextual information. These spatial-wise attention maps are utilized to calibrate features from DESA $\boldsymbol{X}_{SA}$, enhancing their spatial representations. This spatial-wise attention is implemented by cascading a channel pooling layer, a $7\times7\times7$ convolutional layer, and a Sigmoid function
\begin{align}
    \boldsymbol{X}'_{SA}&=\boldsymbol{X}_{SA}*\textrm{Sigmoid}(\textrm{Conv7}(\textrm{ChPool}(\boldsymbol{X}_{LK}))).
\end{align}
In contrast, DESA captures channel-wise global information, while MHLK lacks abilities to capture this information due to the limitation of convolutional kernels. Thus, channel-wise attention maps are extracted from output features of DESA $\boldsymbol{X}_{SA}$ to model inter-dependencies among channels. These channel-wise attention maps are utilized to calibrate features from MHLK $\boldsymbol{X}_{LK}$, improving them with channel-wise global information. This channel-wise attention is implemented by cascading an adaptive average pooling layer, a linear layer, and a Sigmoid function.
\begin{align}
    \boldsymbol{X}'_{LK}=\boldsymbol{X}_{LK}*\textrm{Sigmoid}(\textrm{Linear}(\textrm{AvgPool}(\boldsymbol{X}_{SA})))
\end{align}
Finally, these two calibrated features are fused via addition as the output features $\boldsymbol{X}_{out}\in\mathbb{R}^{C\times D\times H\times W}$. A residual connection is applied
\begin{equation} 
\begin{aligned}
    \boldsymbol{X}_{out}&=\boldsymbol{X}'_{LK}+\boldsymbol{X}'_{SA}, \\
    \boldsymbol{X}_{out}&=\boldsymbol{X}_{out}+\boldsymbol{X}_{in}.
\end{aligned}
\end{equation}

\subsection{Attention-gated Channel MLP module}
We propose an Attention-gated Channel MLP (AG-MLP) module by introducing a spatial-wise attention gate into the channel MLP module (Figure \ref{fig2}F). Specifically, given the input features $\boldsymbol{X}_{in}\in\mathbb{R}^{C\times D\times H\times W}$, a linear projection layer is employed to expand channels by 4 as features $\boldsymbol{X}\in\mathbb{R}^{4C\times D\times H\times W}$ and a GELU function is applied to introduce non-linearity
\begin{align}
    \boldsymbol{X}&=\textrm{GELU}(\textrm{Projection}(\boldsymbol{X}_{in})).
\end{align}
Subsequently, a $3\times3\times3$ depth-wise convolutional layer is employed to extract local spatial features $\boldsymbol{X}_{sp}\in\mathbb{R}^{4C\times D\times H\times W}$
\begin{align}
    \boldsymbol{X}_{sp}=\textrm{DWConv}(\boldsymbol{X}).
\end{align}
Additionally, a spatial-wise attention gate employs a $1\times1\times1$ convolutional layer ($\textrm{Conv}_1$) to capture spatial-wise global information and a Sigmoid function to convert them to attention-gating scores $\boldsymbol{A}\in\mathbb{R}^{4C\times D\times H\times W}$
\begin{align}
    \boldsymbol{A}=\textrm{Sigmoid}(\textrm{Conv}_1(\boldsymbol{X})).
\end{align}
Subsequently, features $\boldsymbol{X}_{sp}$ are adjusted by these attention-gating scores to improve their capabilities of learning spatial representations
\begin{align}
    \boldsymbol{X}=\boldsymbol{X}_{sp}*\boldsymbol{A}.
\end{align}
A second linear projection layer is employed to compress channels to the original dimension as $\boldsymbol{X}_{out}\in\mathbb{R}^{C\times D\times H\times W}$. Two drop-out layers are employed before and after it
\begin{align}
    \boldsymbol{X}_{out}=\textrm{Dropout}(\textrm{Projection}(\textrm{Dropout}(\boldsymbol{X}))).
\end{align}

\subsection{Entangled Transformer Large Kernel block}
\begin{figure*}[!t]
\centering
\includegraphics[width=\textwidth]{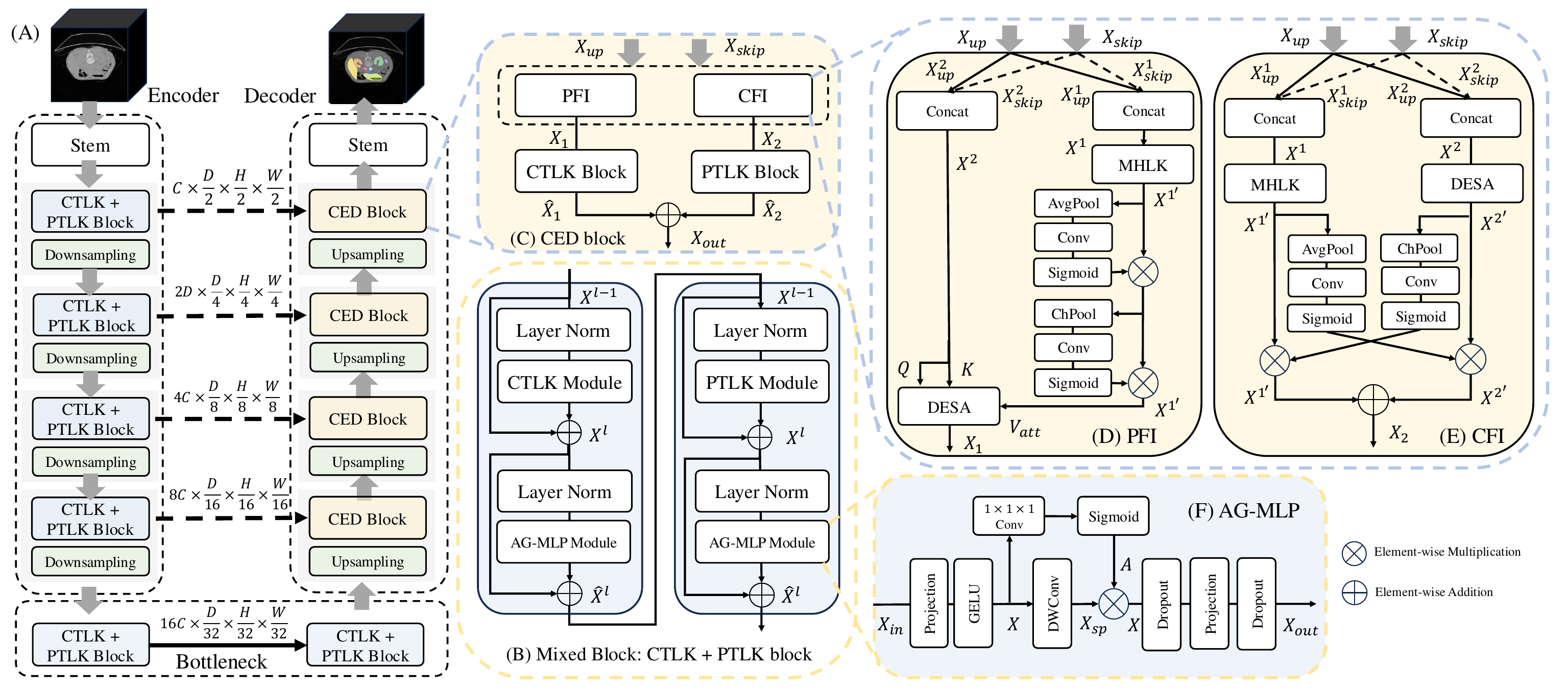}
\caption{The architecture of (A) TransLK-Net, (B) mixed block (CTLK+PTLK block), (C) CED block, (D) PFI module, (E) CFI module, and (F) AG-MLP module. (A) TransLK-Net consists of an encoder, a bottleneck, and a decoder. The encoder and the bottleneck employ a mixed block at each stage, and the decoder employs a CED block at each stage. (B) The mixed block is built by stacking a CTLK and PTLK block. Each CTLK and PTLK block employs a CTLK and PTLK module, respectively, and both of them employ an AG-MLP module. (C) The CED block employs two parallel paths to fuse features $X_{up}$ and $X_{skip}$ and generate $X_1$ and $X_2$ in the PFI and CFI module. Each path employs a CTLK and PTLK block to generate features $\hat{X}_1$ and $\hat{X}_2$, and they are added as output $X_{out}$. (D) the PFI module splits upsampled features $X_{up}$ to $X_{up}^1$ and $X_{up}^2$ and skip-connected features $X_{skip}$ to $X_{skip}^1$ and $X_{skip}^2$. These sub-features are cross-grouped as $X^1$ and $X^2$. Subsequently, features $\boldsymbol{X}^{1'}$ are extracted from $X^1$ in MHLK and enhanced with Progressive Attention Entanglement. DESA takes $X_2$ and $\boldsymbol{X}^{1'}$ as input to generate output features $X_1$. (E) the CFI module employs the same strategy to generate $X^1$ and $X^2$ from $X_{up}$ and $X_{skip}$. Features $X^{1'}$ and $X^{2'}$ are extracted from $X^1$ and $X^2$ in MHLK and DESA, and subsequently enhanced with Collaborative Attention Entanglement as output $X_2$. (F) the AG-MLP module projects input features $X_{in}$ to $X$, and employs a depth-wise convolutional layer to capture spatial-wise features $X_{sp}$. Attention-gating scores $A$ are extracted from $X$, and used to calibrate spatial features $X_{sp}$. Features are projected back to output $X_{out}$.} 
\label{fig2}
\end{figure*}

The Progressively Entangled Transformer Large Kernel block (PTLK$_B$) is constructed by using the PTLK module and AG-MLP to replace self-attention and MLP in a standard hierarchical ViT block, respectively (Figure \ref{fig2}B). Two Layer Normalization (LN) layers are employed before PTLK and AG-MLP. A residual connection is also applied to each module. The output of the $l$-th layer in PTLK$_B$ can be computed as
\begin{equation} 
\begin{aligned}
    \boldsymbol{X}^l&=\textrm{PTLK}(\textrm{LN}(\boldsymbol{X}^{l-1})) + \boldsymbol{X}^{l-1},\\
    \hat{\boldsymbol{X}}^l&=\textrm{AG-MLP}(\textrm{LN}(\boldsymbol{X}^l)) + \boldsymbol{X}^l.
\end{aligned}
\end{equation}
The Collaboratively Entangled Transformer Large Kernel block (CTLK$_B$) is constructed in the same way as PTLK$_B$ (Figure \ref{fig2}B). The output of the $l$-th layer in CTLK$_B$ can be computed as
\begin{equation} 
\begin{aligned}
    \boldsymbol{X}^l&=\textrm{CTLK}(\textrm{LN}(\boldsymbol{X}^{l-1})) + \boldsymbol{X}^{l-1},\\
    \hat{\boldsymbol{X}}^l&=\textrm{AG-MLP}(\textrm{LN}(\boldsymbol{X}^l)) + \boldsymbol{X}^l.
\end{aligned}
\end{equation}

\subsection{Cross Entanglement Decoding block}
The CED block employs a dual-path architecture (Figure \ref{fig2}C). Specifically, the first path employs a Progressive Feature Interaction (PFI) module to fuse skip-connected features from the encoder $\boldsymbol{X}_{skip}\in\mathbb{R}^{C\times D\times H\times W}$ and upsampled features in the decoder $\boldsymbol{X}_{up}\in\mathbb{R}^{C\times D\times H\times W}$, generating features $\boldsymbol{X}_1\in\mathbb{R}^{C\times D\times H\times W}$. Subsequently, a CTLK$_B$ is applied to decode more contextual information from $\boldsymbol{X}_1\in\mathbb{R}^{C\times D\times H\times W}$ as $\hat{\boldsymbol{X}}_1\in\mathbb{R}^{C\times D\times H\times W}$
\begin{equation} 
\begin{aligned}
    \boldsymbol{X}_1 &= \textrm{PFI}(\boldsymbol{X}_{skip};\boldsymbol{X}_{up}),\\
    \hat{\boldsymbol{X}}_1&= \textrm{CTLK}_B (\boldsymbol{X}_1).
\end{aligned}
\end{equation}
Another path employs a Collaborative Feature Interaction (CFI) module to fuse features $\boldsymbol{X}_{skip}$ and $\boldsymbol{X}_{up}$, generating features $\boldsymbol{X}_2\in\mathbb{R}^{C\times D\times H\times W}$. Subsequently, a PTLK$_B$ is applied to decode more contextual information from $\boldsymbol{X}_2\in\mathbb{R}^{C\times D\times H\times W}$ as $\hat{\boldsymbol{X}}_2\in\mathbb{R}^{C\times D\times H\times W}$
\begin{equation} 
\begin{aligned}
    \boldsymbol{X}_2 &= \textrm{CFI}(\boldsymbol{X}_{skip};\boldsymbol{X}_{up}),\\
    \hat{\boldsymbol{X}}_2&= \textrm{PTLK}_B (\boldsymbol{X}_2).
\end{aligned}
\end{equation}
Lastly, output features of the CED block $\boldsymbol{X}_{out}\in\mathbb{R}^{C\times D\times H\times W}$ are generated by adding features $\hat{\boldsymbol{X}}_1$ and $\hat{\boldsymbol{X}}_2$
\begin{align}
    \boldsymbol{X}_{out} = \hat{\boldsymbol{X}}_1 + \hat{\boldsymbol{X}}_2.
\end{align}

\subsubsection{Progressive Feature Interaction module}
In the PFI module (Figure \ref{fig2}D), features $\boldsymbol{X}_{skip}$ are split along channels to generate $\boldsymbol{X}^1_{skip}$ and $\boldsymbol{X}^2_{skip}$, and $\boldsymbol{X}_{up}$ are also split to generate $\boldsymbol{X}^1_{up}$ and $\boldsymbol{X}^2_{up}$ ($\boldsymbol{X}^1_{skip}$,$\boldsymbol{X}^2_{skip}$,$\boldsymbol{X}^1_{up}$,$\boldsymbol{X}^2_{up}\in\mathbb{R}^{\frac{C}{2}\times D\times H\times W}$). These sub-features are cross-grouped as $\boldsymbol{X}^1\in\mathbb{R}^{C\times D\times H\times W}$ and $\boldsymbol{X}^2\in\mathbb{R}^{C\times D\times H\times W}$ via concatenation
\begin{equation} 
\begin{aligned}
    \boldsymbol{X}^1 &= \textrm{Concat}([\boldsymbol{X}^1_{skip}, \boldsymbol{X}^1_{up}]), \\
    \boldsymbol{X}^2 &= \textrm{Concat}([\boldsymbol{X}^2_{skip}, \boldsymbol{X}^2_{up}]).
\end{aligned}
\end{equation}
Subsequently, the interactions between features $\boldsymbol{X}^1$ and $\boldsymbol{X}^2$ are explored using MHLK, DESA, and Progressive Attention Entanglement. Specifically, MHLK is employed to extract finer spatial features $\boldsymbol{X}^{1'}$ from $\boldsymbol{X}^1$, and the Progressive Attention Entanglement is utilized to enhance $\boldsymbol{X}^{1'}$. Lastly, after projecting $\boldsymbol{X}^{1'}$ to $\boldsymbol{V}_{att}$ and $\boldsymbol{X}^2$ to $\boldsymbol{Q}$ and $\boldsymbol{K}$, DESA is applied to model correlations between $\boldsymbol{X}^{1'}$ and $\boldsymbol{X}^2$ to generate the output features $\boldsymbol{X}_1$ of PFI.

\subsubsection{Collaborative Feature Interaction module}
In the CFI module (Figure \ref{fig2}E), we utilize the same strategy to split features $\boldsymbol{X}_{skip}$ and $\boldsymbol{X}_{up}$ and cross-group them to $\boldsymbol{X}^1$ and $\boldsymbol{X}^2$ as PFI. Subsequently, the interactions between features $\boldsymbol{X}^1$ and $\boldsymbol{X}^2$ are explored using MHLK, DESA, and Collaborative Attention Entanglement. Specifically, MHLK is employed to extract finer spatial local features $\boldsymbol{X}^{1'}$ from $\boldsymbol{X}^1$, and DESA is employed to capture global channel features $\boldsymbol{X}^{2'}$ from $\boldsymbol{X}^2$. Then Collaborative Attention Entanglement is employed to improve mutual calibration between $\boldsymbol{X}^{1'}$ and $\boldsymbol{X}^{2'}$ and fuse them, generating output features $\boldsymbol{X}_2$ of CFI.

\subsection{Architecture of TransLK-Net}
The TransLK-Net is designed as an encoder-decoder architecture to learn hierarchical feature representations (Figure \ref{fig2}). Both encoder and decoder comprise four stages, each with downsampling and upsampling rates of 2, 4, 8, and 16, respectively. A bottleneck is utilized to bridge the encoder and the decoder.

\textbf{Encoder.} The stem employs a convolutional layer with a large $7\times7\times7$ kernel and a stride of $2$ to partition the input image with the dimension of $D\times H\times W$ into overlapping feature embedding. Instead of extracting the non-overlapped patches by linear projections, this design allows the model to capture richer local features from input images \citep{wu2021cvt}. Moreover, employing the large kernel provides a large receptive field, thus capturing more informative features and improving segmentation performance. Additionally, this stem projects features to $C$-dimensional vectors ($C=48$) to generate features with the dimensions of $C\times\frac{D}{2}\times\frac{H}{2}\times\frac{W}{2}$. At each stage, a CTLK block and a PTLK block are stacked as a mixed block for generic feature extraction. Subsequently, a $3\times3\times3$ convolution layer with a stride of $2$ is employed for downsampling. It is utilized to increase the dimension of token features and downsample the feature maps by a factor of 2. The number of feature maps at each stage is 96, 192, 384, and 768, respectively.

\textbf{Bottleneck.} The bottleneck stacks a CTLK and PTLK block to extract features without changing their dimensions.

\textbf{Decoder.} A $2\times2\times2$ transposed convolutional layer with a stride of 2 is employed at each stage. It upsamples feature maps from the last stage by decreasing their channel numbers and upscaling their dimensions by a factor of 2. Subsequently, the CED block is employed to fuse upsampled features with skip-connected features from the encoder, decoding richer contextual information. To recover the feature maps to the original dimension $D\times H\times W$, the stem employs a $2\times2\times2$ transposed convolutional layer with a stride of 2. Lastly, a $1\times1\times1$ convolutional layer is used to produce the voxel-wise predictions for segmentation.

\section{Experiments}
\subsection{Datasets}
We implemented our experiments on four segmentation tasks to evaluate the superiority of our segmentation networks. These tasks differ in image modalities, segmentation complexity, number of structures to be segmented, and spatial and phenotypic heterogeneity (Table \ref{tab1}). They are designed to underline the potential of TransLK-Net to generalize across different segmentation tasks. 

\textbf{Abdominal Multi-organ Segmentation.} We implemented abdominal multi-organ segmentation on the MICCAI 2022 AMOS Challenge dataset \citep{ji2022amos}. It consists of 300 abdominal CT images with voxel-level annotations of 15 organs (Spleen, Right kidney, Left kidney, Gall bladder, Esophagus, Liver, Stomach, Aorta, Postcava, Pancreas, Right Adrenal Gland, Left Adrenal Gland, Duodenum, Bladder, and Prostate). Each CT volume consists of $67\thicksim369$ slices of $512\times512$ pixels with a slice spacing of $1.25\thicksim5.00$ mm.

\textbf{Brain Tumor Segmentation.} The Brain Tumor segmentation dataset is from the Medical Segmentation Decathlon (MSD) Challenge \citep{antonelli2022medical}. It comprises 484 multi-parametric Magnetic Resonance Imaging (MRI) scans with three foreground segmentation labels, including Edema (ED), Enhancing Tumor (ET), and Non-Enhancing Tumor (NET). Four modalities are available for each case: Native T1-weighted image (T1w), post-contrast T1-weighted (T1Gd), T2-weighted (T2w), and T2 Fluid Attenuated Inversion Recovery (T2-FLAIR).

\textbf{Hepatic Vessel Tumor Segmentation.} The MSD Hepatic Vessel Tumor segmentation dataset consists of 303 CT scans with manual annotations \citep{antonelli2022medical}. The target segmentation regions are the hepatic vessels (Vessel) and tumors within the liver (Tumor). They are obtained from patients with a variety of primary and metastatic liver tumors.

\textbf{Abdomen CT Organ Segmentation.}  The fourth dataset is from The Fast and Low GPU memory Abdominal oRgan sEgmentation (FLARE) challenge \citep{ma2022fast}. It consists of 361 CT images with voxel-wise annotations of four abdominal organs, including the liver, the kidneys, the spleen, and the pancreas. It demonstrates a large diversity across various centers, vendors, phases, and diseases.

\subsection{Implementation details}
The experiments were implemented using PyTorch. Models were trained and validated on NVIDIA Tesla A100 PCI-E Passive Single GPU each with 40GB of GDDR5 memory. We used a joint loss function ($\mathcal{L}_{Total}$) that consists of cross-entropy loss ($\mathcal{L}_{CE}$) and dice loss ($\mathcal{L}_{Dice}$)
\begin{align}
    \mathcal{L}_{Total}=\mathcal{L}_{CE}(\hat{y},y)+\mathcal{L}_{Dice}(\hat{y},y)
\end{align}
where $y$ and $\hat{y}$ denote the ground truth annotations and predictions. Implementation details were customized for different segmentation tasks (Table \ref{tab1}).

\begin{table}[!t]
\centering
\caption{Dataset information and implementation details.}
\label{tab1}
\resizebox{0.48\textwidth}{!}{
\begin{tabular}{c|c|c|c|c}
\toprule
Dataset & AMOS 2022 & Brain Tumor & Hepatic Vessel & FLARE \\
\midrule
Tasks & Multi-Organs & Tissues & Tubular$\&$Tumor & Multi-Organs \\
Modalities   & CT & MR & CT & CT \\
Dimensions   & 3D & 3D & 3D & 3D \\
Class Number & 15 & 3 & 2 & 4 \\
\midrule
Batch size & 2 & 2 & 2 & 2 \\
Patch size & $96\times160\times160$ & $128\times128\times128$ & $64\times192\times192$ & $96\times192\times160$ \\
\midrule
Preprocessing & \multicolumn{4}{c}{[5$\%$, 95$\%$] clipping. Z score normalization}\\
\midrule
Epochs        & \multicolumn{4}{c}{1500}  \\
Optimizer     & \multicolumn{4}{c}{AdamW}    \\
lr schedule   & \multicolumn{4}{c}{cosine} \\
\midrule
Initial lr (TransLK-Net) & 5e-3   & 5e-5  & 1e-3   & 1e-3   \\
Decay rate (TransLK-Net) & 3e-5   & 3e-6  & 3e-5   & 3e-5   \\
\midrule
\multirow{6}{*}{Augmentation} & \multicolumn{4}{c}{Rotation: (-30, 30) with 0.2 probability} \\
                  & \multicolumn{4}{c}{Scaling: (0.7, 1.4) with 0.2 probability} \\
                  & \multicolumn{4}{c}{Mirroring: 0.5 probability} \\
                  & \multicolumn{4}{c}{Gaussian noise, Gaussian blurring: 0.15 probability} \\
                  & \multicolumn{4}{c}{Brightness and Contrast: 0.15 probability} \\
                  & \multicolumn{4}{c}{Gamma transformation: 0.15 probability} \\
\bottomrule
\end{tabular}}
\end{table}

We employed the 5-fold cross-validation to generate reliable evaluation results on all datasets. Segmentation performance was evaluated by the Dice Similarity Coefficient (DSC). The architecture complexity was evaluated by the number of parameters (Params), and the computational complexity was evaluated by the number of Floating Point Operations (FLOPs). The Wilcoxon signed-rank test was implemented to statistically quantify the differences in segmentation performance between TransLK-Net and SOTA methods.

\begin{table*}[!t]
\centering
\caption{Comparison of segmentation performance among TransLK-Net and SOTA methods on the 2022 AMOS Abdominal Multi-organ segmentation task. The performance was evaluated using the DSC ($\%$; Mean $\pm$ Standard Deviation). \textbf{Bold} represents the best results. ($^*$: $p<0.01$ with Wilcoxon signed-rank test between TransLK-Net and SOTA methods.)}
\label{tab2}
\resizebox{\textwidth}{!}{
\begin{tabular}{c|c|c|c|c|c|c|c|c|c|c|c|c|c}
\toprule
Tasks &  VNet & nnU-Net & Att U-Net & TransBTS & UNETR & Swin UNETR & nnFormer & SegFormer & UX-Net & MedNeXt & VSmTrans & MixUNETR & TransLK-Net \\
\midrule
Spleen       & 95.07$\pm$8.49  & 96.14$\pm$8.29  & 96.06$\pm$8.26  & 95.81$\pm$8.28  & 92.34$\pm$10.65 & 95.77$\pm$8.38  & 95.18$\pm$8.28 
             & 91.82$\pm$10.54 & 95.98$\pm$8.34  & 95.82$\pm$8.30  & 95.84$\pm$8.32  & 95.04$\pm$8.81  & \textbf{97.58}$\pm$2.01 \\
R. kidney    & 94.82$\pm$7.60  & 95.54$\pm$7.99  & 95.79$\pm$5.81  & 95.51$\pm$5.74  & 92.10$\pm$10.95 & 95.19$\pm$8.25  & 94.02$\pm$6.05 
             & 92.31$\pm$6.77  & 95.28$\pm$8.32  & 95.48$\pm$5.79  & 95.44$\pm$6.18  & 94.41$\pm$9.22  & \textbf{96.86}$\pm$1.38 \\
L. kidney    & 94.69$\pm$8.58  & 95.44$\pm$9.38  & 95.47$\pm$8.79  & 94.85$\pm$10.18 & 91.33$\pm$13.52 & 95.00$\pm$9.88  & 93.79$\pm$9.59 
             & 91.30$\pm$11.44 & 94.92$\pm$10.63 & 94.92$\pm$8.64  & 94.65$\pm$11.38 & 94.29$\pm$10.05 & \textbf{96.64}$\pm$7.54\\
Gall bladder & 76.78$\pm$25.79 & 81.62$\pm$23.68 & 82.07$\pm$23.50 & 81.83$\pm$21.95 & 67.46$\pm$27.89 & 79.94$\pm$25.10 & 81.23$\pm$21.06 
             & 69.55$\pm$26.09 & 80.36$\pm$24.63 & 79.53$\pm$25.35 & 81.62$\pm$23.71 & 78.36$\pm$24.90 & \textbf{84.96}$\pm$21.38 \\
Esophagus    & 79.29$\pm$10.88 & 83.72$\pm$9.28  & 83.44$\pm$9.77  & 81.95$\pm$10.04 & 74.32$\pm$13.68 & 82.75$\pm$9.91  & 78.74$\pm$11.30 
             & 68.29$\pm$13.31 & 83.14$\pm$9.97  & 81.41$\pm$11.14 & 83.24$\pm$9.49  & 79.69$\pm$10.69 & \textbf{86.24}$\pm$8.64 \\
Liver        & 96.66$\pm$2.30  & 97.44$\pm$1.82  & 97.37$\pm$1.98  & 97.20$\pm$2.06  & 95.04$\pm$4.20  & 97.14$\pm$2.38  & 96.73$\pm$2.59 
             & 95.41$\pm$3.22  & 97.20$\pm$2.55  & 97.16$\pm$2.29  & 97.25$\pm$2.19  & 96.65$\pm$3.17  & \textbf{97.58}$\pm$1.80 \\
Stomach      & 87.09$\pm$16.32 & 90.38$\pm$15.37 & 90.23$\pm$15.59 & 89.25$\pm$15.73 & 79.17$\pm$18.70 & 88.53$\pm$16.68 & 88.95$\pm$15.35 
             & 82.36$\pm$16.92 & 88.72$\pm$16.88 & 88.99$\pm$16.20 & 89.85$\pm$15.66 & 87.44$\pm$16.78 & \textbf{92.50}$\pm$15.02\\
Aorta        & 91.10$\pm$6.74  & 93.74$\pm$5.15  & 93.53$\pm$5.37  & 92.85$\pm$5.82  & 89.89$\pm$6.27  & 93.89$\pm$5.32  & 90.85$\pm$7.51 
             & 89.89$\pm$4.38  & 93.77$\pm$5.30  & 90.39$\pm$7.43  & 94.92$\pm$3.21  & 94.13$\pm$3.87  & \textbf{96.05}$\pm$2.58 \\
Postcava     & 85.42$\pm$8.93  & 89.34$\pm$6.59  & 89.02$\pm$6.98  & 87.17$\pm$9.17  & 81.93$\pm$8.24  & 88.95$\pm$6.35  & 83.13$\pm$13.30 
             & 82.80$\pm$6.98  & 88.95$\pm$6.84  & 86.00$\pm$11.02 & 89.99$\pm$5.63  & 88.53$\pm$5.95  & \textbf{91.64}$\pm$5.81 \\
Pancreas     & 80.37$\pm$12.61 & 84.20$\pm$11.81 & 84.22$\pm$12.01 & 82.13$\pm$12.74 & 73.38$\pm$16.00 & 82.74$\pm$13.46 & 78.73$\pm$14.28 
             & 74.32$\pm$14.29 & 83.54$\pm$13.11 & 82.50$\pm$13.28 & 84.13$\pm$11.92 & 81.48$\pm$13.53 & \textbf{86.97}$\pm$8.74\\
R.A. gland   & 72.44$\pm$13.03 & 74.81$\pm$14.45 & 74.69$\pm$14.02 & 73.36$\pm$13.92 & 68.22$\pm$14.94 & 75.32$\pm$13.04 & 69.08$\pm$14.00 
             & 58.38$\pm$11.51 & 75.69$\pm$13.48 & 74.12$\pm$14.29 & 74.99$\pm$12.57 & 72.87$\pm$12.58 & \textbf{79.12}$\pm$7.56\\
L.A. gland   & 73.35$\pm$14.66 & 75.20$\pm$14.97 & 75.43$\pm$14.68 & 73.83$\pm$15.46 & 65.68$\pm$18.02 & 75.30$\pm$15.09 & 69.62$\pm$14.20 
             & 54.59$\pm$14.36 & 75.84$\pm$14.89 & 74.59$\pm$14.94 & 75.81$\pm$14.21 & 72.59$\pm$15.05 & \textbf{80.96}$\pm$7.88\\
Duodenum     & 73.42$\pm$14.97 & 78.94$\pm$14.54 & 78.72$\pm$14.81 & 76.46$\pm$16.11 & 65.05$\pm$15.58 & 77.45$\pm$14.79 & 72.40$\pm$16.41 
             & 67.19$\pm$14.19 & 77.92$\pm$15.27 & 76.48$\pm$15.95 & 79.52$\pm$14.31 & 75.52$\pm$14.86 & \textbf{82.94}$\pm$10.24\\
Bladder      & 83.32$\pm$19.07 & 87.68$\pm$15.95 & 87.89$\pm$15.50 & 87.31$\pm$14.64 & 75.94$\pm$21.69 & 86.75$\pm$16.10 & 85.74$\pm$15.41 
             & 77.83$\pm$18.46 & 86.83$\pm$17.74 & 86.59$\pm$16.01 & 87.51$\pm$15.45 & 84.36$\pm$18.20 & \textbf{90.48}$\pm$12.28\\
Prostate     & 78.30$\pm$19.04 & 82.23$\pm$20.56 & 82.20$\pm$19.48 & 82.12$\pm$18.72 & 73.71$\pm$21.09 & 81.04$\pm$19.67 & 80.74$\pm$18.64 
             & 74.67$\pm$19.51 & 81.13$\pm$20.84 & 80.92$\pm$19.49 & 80.95$\pm$20.86 & 78.96$\pm$19.78 & \textbf{86.32}$\pm$18.08\\
\midrule
Mean         & 84.14$\pm$16.18 & 87.10$\pm$15.16 & 87.08$\pm$14.95 & 86.11$\pm$15.23 & 79.04$\pm$18.99 & 86.38$\pm$15.48 & 83.93$\pm$16.09 & 78.05$\pm$18.74 & 86.62$\pm$15.65 & 85.66$\pm$15.82 & 87.05$\pm$14.96 & 84.96$\pm$16.06 & \textbf{89.79}$^*\pm$11.46 \\
\bottomrule
\end{tabular}}
\end{table*}

\begin{table*}[!t]
\centering
\caption{Comparison of segmentation performance among TransLK-Net and SOTA methods on the Brain Tumor segmentation, Hepatic Vessel Tumor segmentation, and FLARE Abdomen Organ segmentation. The performance was evaluated using the DSC ($\%$; Mean $\pm$ Standard Deviation). \textbf{Bold} represents the best results. ($^*$: $p<0.01$ with Wilcoxon signed-rank test between TransLK-Net and SOTA methods.)}
\label{tab3}
\resizebox{\textwidth}{!}{
\begin{tabular}{c|c|ccc|c|cc|c|cccc}
\toprule
Tasks & \multicolumn{4}{c|}{Brain Tumor Segmentation} & \multicolumn{3}{c|}{Hepatic Vessel Tumor Segmentation} & \multicolumn{5}{c}{FLARE Abdomen Organ segmentation} \\
\midrule
Methods    &  Mean & ET & ED & NET & Mean & Vessel & Tumor & Mean & Liver & Kidney & Spleen & Pancreas \\
\midrule
V-Net      & 72.55$\pm$21.97 & 77.08$\pm$22.96 & 79.33$\pm$13.08 & 61.28$\pm$23.59 & 64.30$\pm$21.76 & 62.44$\pm$12.41 & 66.16$\pm$28.03 
           & 93.54$\pm$8.40  & 98.24$\pm$1.07  & 96.07$\pm$3.50  & 98.04$\pm$1.40  & 81.84$\pm$9.00 \\
nnU-Net    & 73.50$\pm$21.44 & 78.79$\pm$21.48 & 80.33$\pm$12.23 & 61.37$\pm$23.38 & 65.38$\pm$21.16 & 63.45$\pm$12.54 & 67.31$\pm$27.04 
           & 94.38$\pm$7.50  & 98.53$\pm$0.81  & 96.61$\pm$3.22  & 98.29$\pm$1.04  & 84.11$\pm$8.36 \\
Att U-Net  & 73.43$\pm$21.48 & 78.56$\pm$21.69 & 80.38$\pm$11.97 & 61.35$\pm$23.45 & 65.91$\pm$21.20 & 62.77$\pm$12.54 & 69.05$\pm$26.87 
           & 94.43$\pm$7.37  & 98.53$\pm$0.77  & 96.63$\pm$3.02  & 98.26$\pm$0.93  & 84.32$\pm$8.25 \\
TransBTS   & 73.80$\pm$21.17 & 78.87$\pm$21.64 & 80.78$\pm$11.60 & 61.74$\pm$22.84 & 64.94$\pm$21.11 & 61.76$\pm$12.17 & 68.13$\pm$26.89 
           & 93.96$\pm$8.32  & 98.44$\pm$1.03  & 96.43$\pm$3.71  & 98.19$\pm$1.84  & 82.78$\pm$9.45 \\
UNETR      & 72.56$\pm$21.70 & 78.00$\pm$22.04 & 79.73$\pm$11.65 & 59.94$\pm$23.48 & 57.24$\pm$24.49 & 61.80$\pm$12.36 & 52.67$\pm$31.70 
           & 92.19$\pm$10.61 & 97.87$\pm$1.78  & 95.99$\pm$3.25  & 96.95$\pm$6.79  & 77.96$\pm$10.89 \\
Swin UNETR & 73.52$\pm$21.36 & 78.53$\pm$21.72 & 80.49$\pm$11.92 & 61.55$\pm$23.20 & 62.63$\pm$22.62 & 61.95$\pm$12.61 & 63.32$\pm$29.39 
           & 93.78$\pm$8.48  & 98.35$\pm$1.56  & 96.47$\pm$3.52  & 97.95$\pm$3.70  & 82.36$\pm$9.13 \\
nnFormer   & 73.52$\pm$21.31 & 78.64$\pm$21.57 & 80.37$\pm$12.06 & 61.60$\pm$23.16 & 66.06$\pm$20.30 & 62.61$\pm$12.10 & 69.51$\pm$25.57 
           & 93.81$\pm$7.77  & 98.29$\pm$1.00  & 95.97$\pm$2.35  & 97.93$\pm$1.46  & 83.05$\pm$8.67 \\
SegFormer  & 46.00$\pm$26.41 & 40.20$\pm$26.66 & 63.84$\pm$17.25 & 33.95$\pm$24.26 & 58.73$\pm$21.97 & 54.37$\pm$10.32 & 63.10$\pm$28.64 
           & 92.58$\pm$8.99  & 97.90$\pm$0.98  & 95.23$\pm$3.14  & 96.98$\pm$5.29  & 80.22$\pm$8.77 \\
UX Net     & 60.86$\pm$22.78 & 64.90$\pm$24.74 & 66.65$\pm$15.37 & 51.04$\pm$23.75 & 63.49$\pm$21.65 & 62.65$\pm$12.41 & 64.33$\pm$27.96 
           & 93.84$\pm$8.54  & 98.41$\pm$1.29  & 96.46$\pm$3.94  & 98.21$\pm$1.42  & 82.30$\pm$9.59 \\
MedNeXt    & 73.86$\pm$21.29 & 79.14$\pm$21.32 & 80.70$\pm$11.64 & 61.73$\pm$23.40 & 66.01$\pm$20.61 & 62.54$\pm$12.58 & 69.49$\pm$25.82 
           & 94.11$\pm$7.75  & 98.40$\pm$1.11  & 96.44$\pm$3.50  & 98.17$\pm$1.28  & 83.45$\pm$8.42 \\
VSmTrans   & 73.98$\pm$21.30 & 79.22$\pm$21.47 & 80.77$\pm$11.83 & 61.94$\pm$23.28 & 64.88$\pm$20.60 & 62.05$\pm$12.81 & 67.71$\pm$25.86 
           & 94.18$\pm$7.62  & 98.41$\pm$1.20  & 96.49$\pm$3.50  & 98.20$\pm$1.59  & 83.65$\pm$8.14 \\
MixUNETR   & 74.08$\pm$20.85 & 79.06$\pm$21.75 & 80.67$\pm$11.53 & 62.55$\pm$22.30 & 62.64$\pm$21.42 & 62.47$\pm$12.29 & 62.81$\pm$27.69 
           & 93.94$\pm$8.03  & 98.37$\pm$1.19  & 96.49$\pm$3.35  & 98.08$\pm$1.92  & 82.82$\pm$8.63 \\
\midrule
TransLK-Net  & \textbf{74.73}$^*\pm$19.86 & \textbf{80.13}$\pm$20.89 & \textbf{81.42}$\pm$9.31 & \textbf{62.64}$\pm$22.18 
             & \textbf{67.89}$^*\pm$20.12 & \textbf{64.91}$\pm$10.66 & \textbf{70.87}$\pm$26.04 & \textbf{94.83}$^*\pm$6.04 & \textbf{98.60}$\pm$1.03 & \textbf{97.17}$\pm$1.27 & \textbf{98.33}$\pm$0.88 & \textbf{85.22}$\pm$5.17 \\
\bottomrule
\end{tabular}}
\end{table*}

\begin{table}[!t]
\centering
\caption{Comparison of model complexity among TransLK-Net and SOTA methods. Params and FLOPs were evaluated using input patches with dimensions of $96\times96\times96$.}
\label{tab4}
\begin{tabular}{c|cc}
\toprule
Methods    & Params (M) & FLOPs (G) \\
\midrule
VNet       & 45.66    & 370.52  \\
Att U-Net  & 69.08    & 360.98  \\
nnU-Net    & 68.38    & 357.13  \\
TransBTS   & 31.58    & 110.69  \\
UNETR      & 92.78    & 82.73   \\
Swin UNETR & 62.19    & 329.28  \\
nnFormer   & 149.33   & 284.28  \\
SegFormer  & 4.50     & 5.02    \\
UX Net     & 53.01    & 632.33  \\
MedNeXt    & 11.65    & 178.05  \\
VSmTrans   & 50.39    & 358.21  \\
MixUNETR   & 62.03    & 329.99  \\
\midrule
TransLK-Net  & 40.95 & 86.17 \\
\bottomrule
\end{tabular}
\end{table}

\subsection{Comparison of with State-of-the-arts}
We compared the performance of TransLK-Net with various SOTA 3D volumetric segmentation models for a thoughtful comparison. These methods include
\begin{itemize}
    \item CNN-based methods: VNet \citep{milletari2016v}, nnU-Net \citep{isensee2021nnu}, Attention gated U-Net (Att U-Net) \citep{oktay2018attention}
    \item ViT-based methods: nnFormer \citep{zhou2023nnformer}, SegFormer \citep{perera2024segformer3d}
    \item Hybrid CNN-ViT-based methods: TransBTS \citep{wang2021transbts}, UNETR \citep{hatamizadeh2022unetr}, Swin UNETR \citep{hatamizadeh2021swin}, 3D UX Net \citep{lee2023d}, MedNeXt \citep{roy2023mednext}, VSmTrans \citep{liu2024vsmtrans} and MixUNETR \citep{shen2025mixunetr}
\end{itemize}

\subsubsection{Comparison of TransLK-Net with State-of-the-arts}

\begin{figure*}[!h]
\centering
\includegraphics[width=0.9\textwidth]{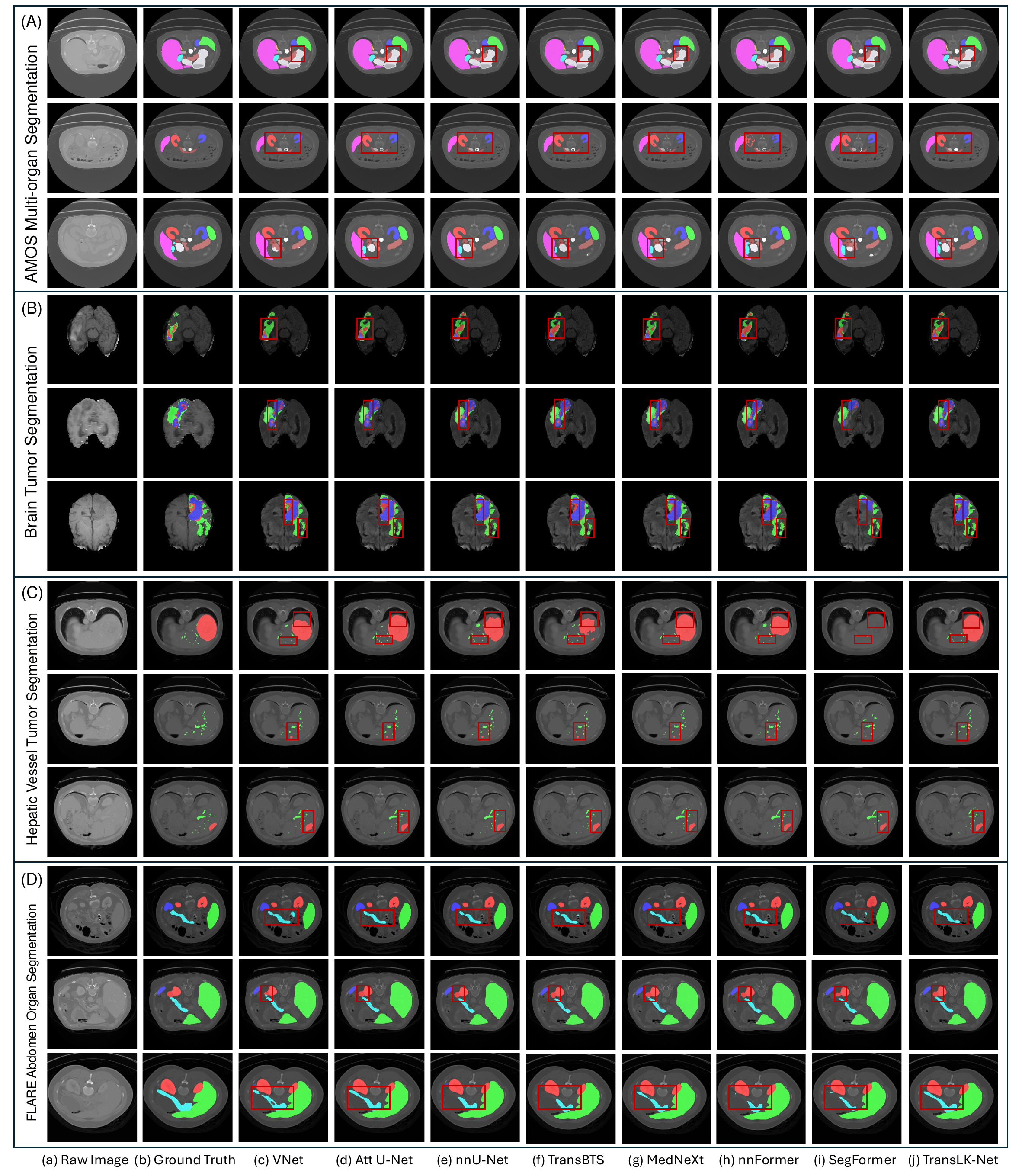}
\caption{Qualitative comparison between (j) TransLK-Net and (a) Raw image, (b) ground truth, (c) VNet, (d) Att U-Net, (e) nnU-Net, (f) TransBTS, (g) MedNeXt, (h) nnFormer, and (i) SegFormer. Results are shown across four public datasets, including (A) the AMOS 2022 Multi-organ dataset, (B) the MSD Brain Tumor dataset, (C) the MSD Hepatic Vessel Tumor dataset, and (D) the FLARE Abdomen Organ dataset. Red boxes mark the regions where TransLK-Net demonstrates higher segmentation quality than other methods.}
\label{vis1}
\end{figure*}

\textbf{AMOS Multi-organ segmentation.} TransLK-Net achieved the best and most robust segmentation performance in multi-organ segmentation from CT, demonstrating the highest mean DSC score of $89.76$ and the lowest standard deviation of $11.68$ than other SOTA methods (Table \ref{tab2} and Figure \ref{vis1}). Specifically, TransLK-Net achieved superior segmentation performance than CNN-based methods while maintaining much lower computational complexity. Att U-Net and nnU-Net achieved better performance than other methods, but TransLK-Net demonstrated $2.69$ and $2.71$ higher points in DSC scores with $69\%$ and $67\%$ fewer Params and $319\%$ and $314\%$ fewer FLOPs, respectively. Additionally, TransLK-Net demonstrated much higher segmentation performance than ViT-based methods. Although SegFormer presented lower complexity, TransLK-Net achieved $11.74$ higher points in DSC score. TransLK-Net outperformed hybrid CNN-ViT methods. VSmTrans demonstrated the highest DSC score among hybrid CNN-ViT methods, but TransLK-Net achieved $2.74$ higher DSC points with $23\%$ fewer Params and $316\%$ fewer FLOPs. Although MedNeXt and TransBTS demonstrated fewer Params, TransLK-Net achieved $4.13$ and $3.68$ higher DSC points with $107\%$ and $28\%$ fewer FLOPs than them. Additionally, TransLK-Net demonstrated much better segmentation performance than SOTA methods in organs with heterogeneous shapes and sizes, such as adrenal glands, pancreas, duodenum, and aorta. Thus, TransLK-Net has superior capabilities of capturing varying-scaled features from multiple organs.

\textbf{Brain Tumor segmentation.} TransLK-Net demonstrated superior segmentation performance than SOTA methods in brain tumor segmentation from MR imaging, achieving $74.73$ points in the mean DSC score (Table \ref{tab3} and Figure \ref{vis1}). Specifically, MixUNETR demonstrated better performance than other SOTA methods in this task, but TransLK outperformed it by $0.65$ points in DSC with $51\%$ fewer Params and $283\%$ fewer FLOPs. nnU-Net achieved the better segmentation performance than other CNN-based methods, but TransLK-Net outperformed it by $1.23$ points. nnFormer was initially designed to advance brain tumor segmentation in MR images, but TransLK-Net achieved $1.21$ higher points in DSC with $265\%$ fewer Params and $230\%$ fewer FLOPs. Compared to Swin UNETR which was proposed as a hybrid CNN-ViT method for brain tumor segmentation, but TransLK-Net achieved $1.21$ higher points in DSC with $52\%$ fewer Params and $282\%$ fewer FLOPs.

\textbf{Hepatic Vessel Tumor segmentation.} TransLK-Net achieved the highest DSC scores of $67.89$ with the lowest standard deviation of $20.12$ among various methods, demonstrating superior and robust performance in this hepatic vessel and tumor segmentation task (Table \ref{tab3} and Figure \ref{vis1}). Specifically, compared with Att U-Net which demonstrated better performance than other CNN-based methods, TransLK-Net achieved approximately $2$ higher points in DSC. nnFormer achieved the best performance among SOTA methods, but TransLK-Net outperformed it by $1.83$ DSC points. Among various hybrid CNN-ViT methods, MedNeXt showed the best performance, but TransLK-Net demonstrated $1.88$ higher DSC points. UX Net employs large kernel-based convolutions in hybrid CNN-ViT architecture to capture features by large receptive fields, but TransLK-Net achieved $4.4$ higher points while employing $29\%$ fewer Params and $634\%$ fewer FLOPs. Additionally, TransLK-Net achieved the highest DSC points among various methods in vessel and tumor segmentation tasks. Thus, TransLK-Net presented superior capabilities for capturing features from tubular structures and heterogeneous tissues.

\textbf{FLARE Abdomen Organ segmentation.} In this diverse abdominal organ segmentation task, TransLK-Net demonstrated superior segmentation performance, achieving the highest $94.83$ DSC points and the lowest $6.04$ standard deviation (Table \ref{tab3} and Figure \ref{vis1}). Specifically, Att U-Net and nnU-Net achieved better segmentation results than other SOTA methods, but TransLK-Net demonstrated $0.4$ and $0.45$ higher DSC points than them, respectively. TransLK-Net achieved approximately $1-2.4$ higher DSC scores than ViT-based methods. Additionally, TransLK-Net demonstrated $0.72$ and $0.65$ higher DSC scores than two hybrid CNN-ViT methods, MedNext and VSmTrans.

\subsection{Ablation studies}
We implemented ablation studies on PTLK and CTLK blocks, the CED block, the AG-MLP module, and the optimal head number to investigate their impact on segmentation performance and computational complexity.

We constructed TransLK-ViT from TransLK-Net for easily implementing ablation studies. Instead of the decoder in the TransLK-Net, TransLK-ViT employed a ViT-based decoder which was symmetric to its encoder. Specifically, a CTLK block and a PTLK block were stacked and incorporated into each decoder stage. Features skip-connected from the encoder $\boldsymbol{X}_{skip}$ and features upsampled from the last stage $\boldsymbol{X}_{up}$ were concatenated along channels. Subsequently, a $1\times1\times1$ convolutional layer was employed to recover the channel number. This fusion strategy was commonly utilized in U-shaped encoder-decoder segmentation models.

We implemented TransLK-NETR by incorporating PTLK and CTLK blocks into a widely used hybrid CNN-ViT architecture (Figure \ref{fig3}). This architecture is employed in several volumetric segmentation methods, such as UNETR \citep{hatamizadeh2022unetr}, Swin UNETR \citep{hatamizadeh2021swin}, UX Net \citep{lee2023d}, VSmTrans \citep{liu2024vsmtrans}, and MixUNETR \citep{shen2025mixunetr}. These networks utilize the same CNN-based decoder, but they incorporate different blocks into the ViT-based encoder. In the TransLK-NETR, the encoder stacks a CTLK and PTLK block as a mixed block at each stage. 

\begin{table}[!t]
\centering
\caption{Comparison of segmentation performance and computational complexity among different basic modules in TransLK-ViT and TransLK-NETR on 2022 AMOS Multi-organ segmentation task. Performance was evaluated using DSC ($\%$; Mean$\pm$STD). The Params and FLOPs were evaluated using input patches with dimensions of $96\times96\times96$.}
\label{tab5}
\resizebox{0.48\textwidth}{!}{
\begin{tabular}{c|c|c|cc}
\toprule
Backbones     & Blocks & DSC & Params (M) & FLOPs (G) \\
\midrule
\multirow{5}{*}{TransLK-ViT} & MHLK + MHLK  & 87.78$\pm$14.90 & 34.61 & 66.32 \\
& DESA + DESA                               & 87.93$\pm$14.66 & 35.84 & 71.18 \\
& PTLK + PTLK                               & 88.66$\pm$13.95 & 38.55 & 78.07 \\
& CTLK + CTLK                               & 88.62$\pm$14.17 & 40.51 & 78.05 \\
& CTLK + PTLK                               & 89.06$\pm$13.86 & 39.53 & 78.06 \\
\midrule
\multirow{5}{*}{TransLK-NETR} & MHLK + MHLK & 88.61$\pm$13.74 & 45.29 & 338.77 \\
& DESA + DESA                               & 88.67$\pm$13.66 & 45.45 & 341.15 \\
& PTLK + PTLK                               & 89.04$\pm$13.24 & 46.08 & 344.58 \\
& CTLK + CTLK                               & 88.92$\pm$13.18 & 46.47 & 344.57 \\
& CTLK + PTLK                               & 89.32$\pm$12.46 & 46.28 & 344.57 \\
\bottomrule
\end{tabular}}
\end{table}

\begin{figure}[!h]
\centering
\includegraphics[width=0.48\textwidth]{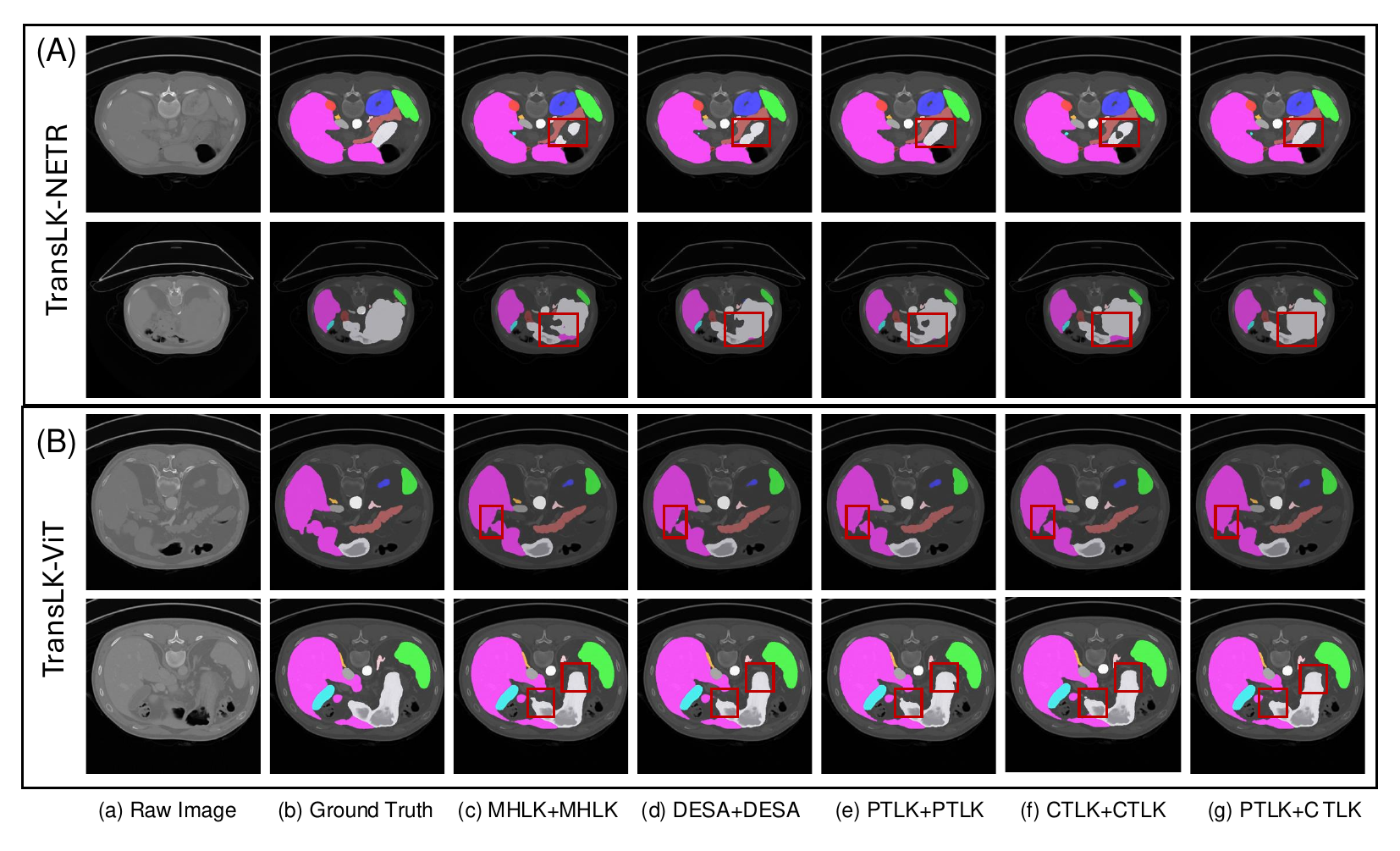}
\caption{Qualitative comparison of (A) TransLK-NETR and (B) TransLK-ViT with their different designs on AMOS multi-organ segmentation task. They were incorporated with (c) two MHLK blocks (MHLK+MHLK), (d) two DESA blocks (DESA+DESA), (e) two PTLK blocks (PTLK+PTLK), (f) two CTLK blocks (CTLK+CTLK), and (g) the mixed blocks (CTLK+PTLK). Red boxes mark the regions where the mixed blocks demonstrate better segmentation results than others.}
\label{vis3}
\end{figure}

\subsubsection{Analysis of Entanglement Transformer Large Kernel block}
To investigate the impact of the CTLK and PTLK blocks, we incorporated different blocks into TransLK-ViT and TransLK-NETR. The mixed blocks (CTLK+PTLK) were originally incorporated into each stage of these networks. Subsequently, two MHLK blocks (MHLK+MHLK), two DESA blocks (DESA+DESA), two PTLK blocks (PTLK+PTLK), or two CTLK blocks (CTLK+CTLK) were incorporated, separately.

TransLK-ViT achieved the highest DSC scores of $89.06$ when the mixed block (CTLK+PTLK) was incorporated into the encoder and the decoder (Table \ref{tab6} and Figure \ref{vis3}). Specifically, incorporating this mixed block into TransLK-ViT achieved superior performance than incorporating two sub-components, two MHLK or DESA blocks, with a slight increase in Params and FLOPs. Additionally, employing the mixed block achieved $0.4$ and $0.44$ higher DSC scores than employing a uniform block, two PTLK or CTLK blocks, respectively, with relatively the same computational complexity. 

When TransLK-NETR was incorporated the mixed block (CTLK+PTLK), it achieved the highest DSC scores of $89.32$ (Table \ref{tab5} and Figure \ref{vis3}). Incorporation of the mixed block into TransLK-NETR outperformed the incorporation of two MHLK or two DESA blocks, and two PTLK or CTLK blocks.

Employing PTLK and CTLK blocks improved the segmentation performance over employing separate large kernel convolution or self-attention modules. Additionally, PTLK and CTLK blocks utilized different mechanisms to entangle features from MHLK and DESA, so employing the mixed block enhanced the network's capabilities to capture diverse features, thus improving segmentation performance compared with employing the uniform block.

\begin{figure*}[!h]
\centering
\includegraphics[width=0.9\textwidth]{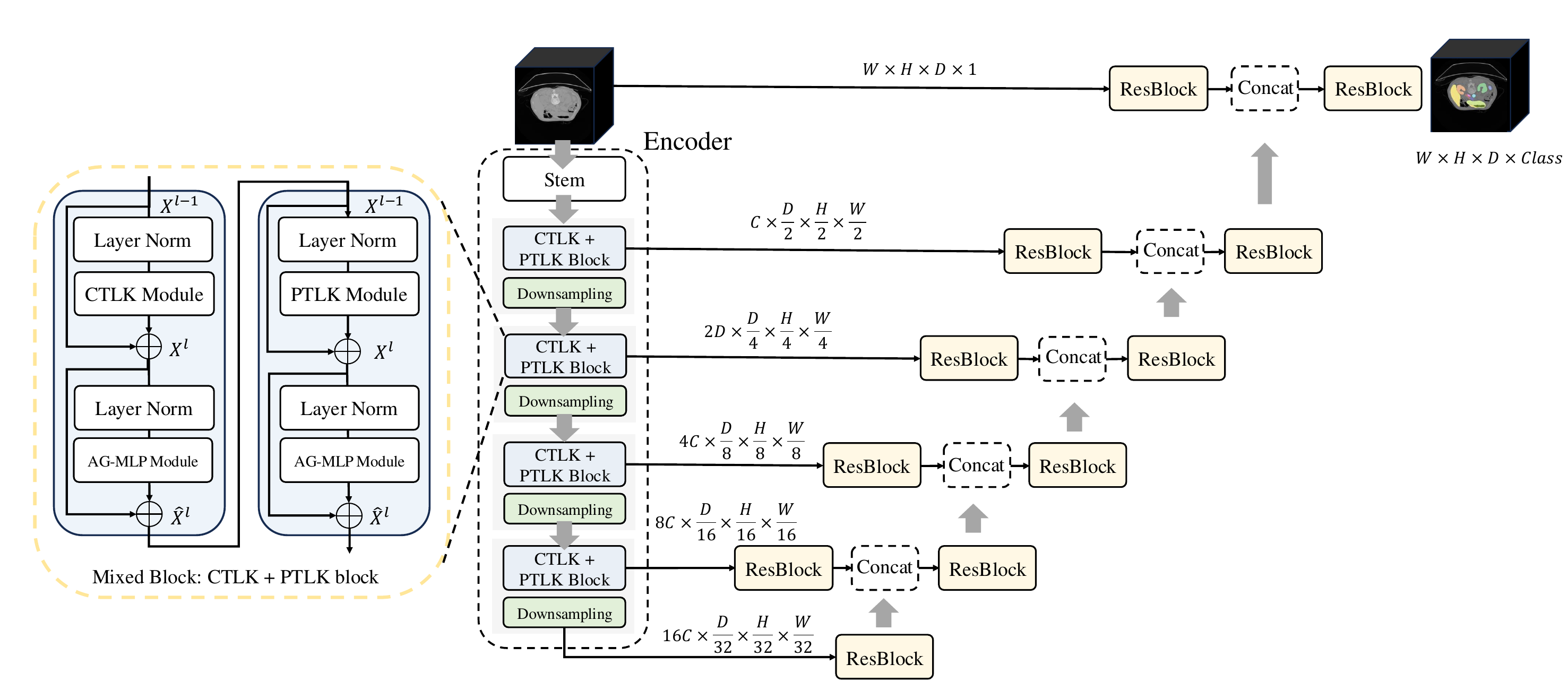}
\caption{The architecture of the TransLK-NETR. The encoder is constructed by incorporating the mixed block (CTLK+PTLK block) into a 4-stage ViT architecture for feature extraction. Each CTLK block consists of a CTLK module and an AG-MLP module, and each PTLK block consists of a PTLK module and an AG-MLP module.} 
\label{fig3}
\end{figure*}

\begin{table*}[!t]
\centering
\caption{Comparison of segmentation performance among TransLK-NETR and other methods on the four tasks. The performance was evaluated using the DSC ($\%$; Mean $\pm$ Standard Deviation). \textbf{Bold} represents the best results. The Params and FLOPs were evaluated using the input patches with the dimension of $96\times96\times96$. ($^*$: $p<0.01$ with Wilcoxon signed-rank test between TransLK-NETR and SOTA methods.)}
\label{tab6}
\resizebox{\textwidth}{!}{
\begin{tabular}{c|c|c|c|c||c|c}
\toprule
Networks     & AMOS 2022 & Brain Tumor & Hepatic Vessel Tumor & FLARE & Params (M) & FLOPs (G) \\
\midrule
UNETR        & 79.04$\pm$18.99 & 72.56$\pm$21.70 & 57.24$\pm$24.49 & 92.19$\pm$10.61 & 92.78 & 82.73 \\
Swin UNETR   & 86.38$\pm$15.48 & 73.52$\pm$21.36 & 62.63$\pm$22.62 & 93.78$\pm$8.48  & 62.19 & 329.28 \\
UX-Net       & 86.62$\pm$15.65 & 60.86$\pm$22.78 & 63.49$\pm$21.65 & 93.84$\pm$8.54  & 53.01 & 632.33 \\
VSmTrans     & 87.05$\pm$14.96 & 73.98$\pm$21.30 & 64.88$\pm$20.60 & 94.18$\pm$7.62  & 50.39 & 358.21 \\
MixUNETR     & 84.96$\pm$16.06 & 74.08$\pm$20.85 & 62.64$\pm$21.42 & 93.94$\pm$8.03  & 62.03 & 329.99 \\
TransLK-NETR & \textbf{89.32}$^*\pm$12.46 & \textbf{74.43}$^*\pm$19.90 & \textbf{67.61}$^*\pm$20.06 & \textbf{94.75}$^*\pm$6.38 & 46.28 & 344.57 \\
\bottomrule
\end{tabular}}
\end{table*}

\subsubsection{Comparison of Entanglement Transformer Large Kernel block with other blocks}
\begin{figure*}[!h]
\centering
\includegraphics[width=0.7\textwidth]{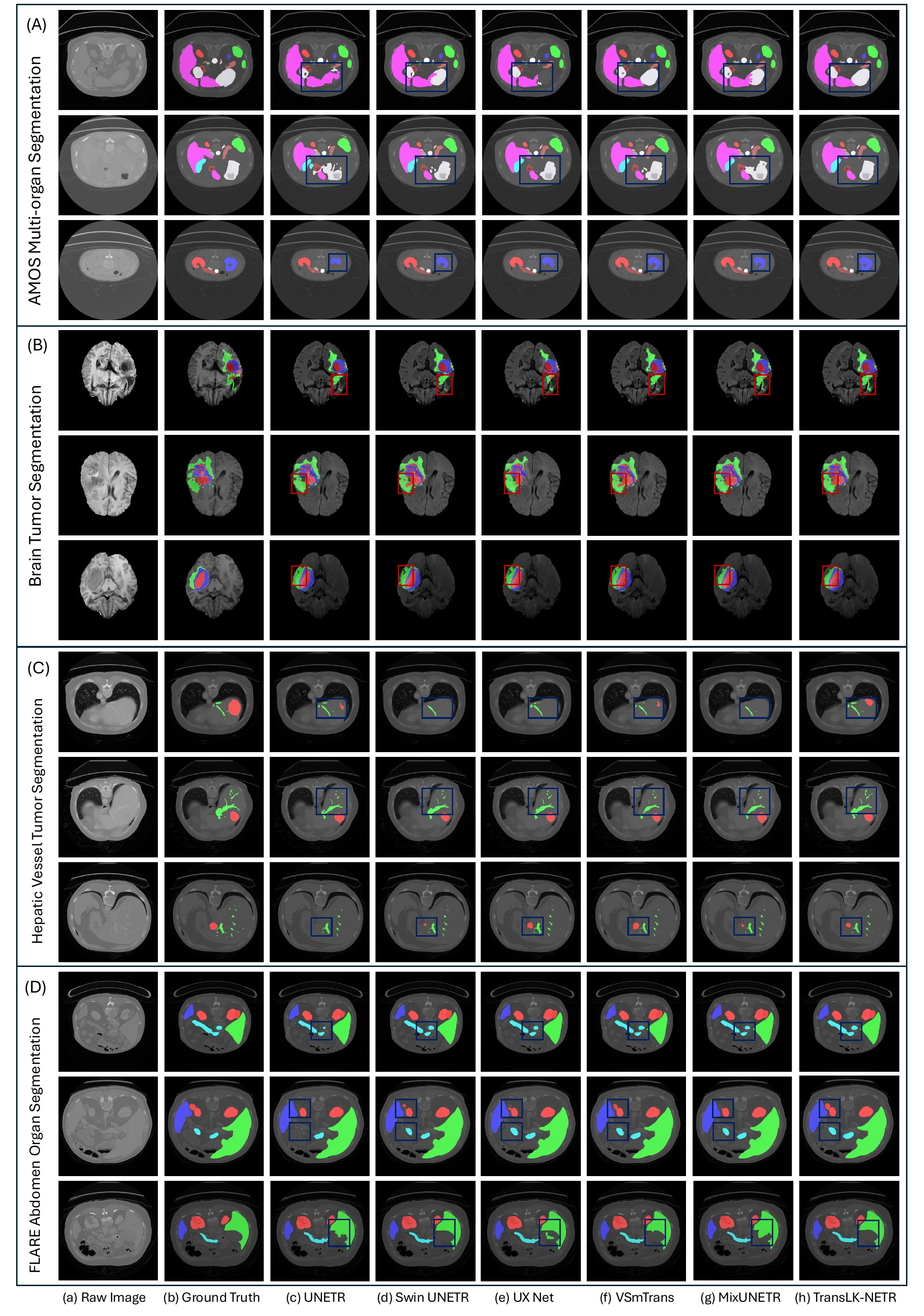}
\caption{Qualitative comparison between (h) TransLK-NETR and (a) Raw image, (b) ground truth, (c) UNETR, (d) Swin UNETR, (e) UX Net, (f) VSmTrans, and (g) MixUNETR. Results are shown across four public datasets, including (A) the AMOS 2022 Multi-organ dataset, (B) the MSD Brain Tumor dataset, (C) the MSD Hepatic Vessel Tumor dataset, and (D) the FLARE Abdomen Organ dataset. Red and blue boxes mark the regions where TransLK-NETR demonstrates better segmentation results than other methods.}
\label{vis2}
\end{figure*}

We compared TransLK-NETR with UNETR, Swin UNETR, UX Net, VSmTrans, and MixUNETR to demonstrate the superior performance of PTLK and CTLK blocks over other blocks. TransLK-NETR demonstrated superior segmentation performance than those hybrid CNN-ViT methods in all four segmentation tasks (Table \ref{tab6} and Figure \ref{vis2}). These results demonstrate the superior capabilities of our PTLK and CTLK blocks over other large kernel and transformer blocks. Specifically, UNETR and Swin UNETR employ ViT and Swin ViT blocks for feature encoding, respectively, but TransLK-NETR achieved $10.28$ and $2.94$ higher points in DSC with $100\%$ and $34\%$ fewer Params than them in AMOS multi-organ segmentation, respectively. UX Net employs large kernel-based convolutions, but TransLK-NETR outperformed it by $2.7$ higher points in DSC while utilizing $15\%$ fewer Params and $84\%$ fewer FLOPs. VSmTrans and MixUNETR employ hybrid convolution-ViT blocks to encode features. TransLK-NETR achieved superior segmentation performance over them by $2.27$ and $4.36$ higher points in DSC. Additionally, TransLK-NETR utilized approximately $9\%$ fewer Params and $4\%$ fewer FLOPs than VSmTrans, and $51\%$ fewer Params than MixUNETR. TransLK-NETR also demonstrated the best performance in the brain tumor segmentation, hepatic vessel tumor segmentation, and abdomen CT organ segmentation tasks.

\subsubsection{Analysis of Cross Entanglement Decoding block}
To evaluate the impact of the CED block, we compared the segmentation performance and computational complexity of TransLK-Net with TransLK-ViT in four tasks. This study was designed to compare the performance of the CED block in TransLK-Net with the common ViT-based decoding block in TransLK-ViT. 

TransLK-Net achieved higher segmentation performance than TransLK-ViT in four tasks, demonstrating the superiority of the CED block over standard fusion strategies in U-shaped methods (Table \ref{tab7}). Specifically, TransLK-Net achieved $0.73$ and $0.39$ higher points in DSC than TransLK-ViT in AMOS and FLARE abdominal multi-organ segmentation tasks, while slightly increasing around $3.6\%$ Params and $10\%$ FLOPs. TransLK-Net demonstrated $0.58$ higher points in DSC than TransLK-ViT in brain tumor segmentation. When segmenting the hepatic vessel and tumor, TransLK-Net and TransLK-ViT achieved competitive performance.

\begin{table}[!t]
\centering
\caption{Comparison of segmentation performance and computational complexity between TransLK-ViT and TransLK-Net on four segmentation tasks. Performance was evaluated using DSC ($\%$; Mean$\pm$STD). The Params and FLOPs were evaluated using input patches with dimensions of $96\times96\times96$.}
\label{tab7}
\resizebox{0.48\textwidth}{!}{
\begin{tabular}{c|c|cc}
\toprule
Tasks     & TransLK-ViT & TransLK-Net \\
\midrule
AMOS Multi-organ    & 89.06$\pm$13.86 & 89.79$\pm$11.46 \\
Brain Tumor         & 74.15$\pm$21.51 & 74.73$\pm$19.86  \\
Vessel Tumor        & 67.80$\pm$19.74 & 67.89$\pm$20.12  \\
FLARE Abdomen Organ & 94.44$\pm$6.87 & 94.83$\pm$6.04  \\
\midrule
Params (M) & 39.53 & 40.95 \\
FLOPs (G)  & 78.06 & 86.17 \\
\bottomrule
\end{tabular}}
\end{table}

\subsubsection{Ablation Study on AG-MLP}
\begin{figure}[!h]
\centering
\includegraphics[width=0.48\textwidth]{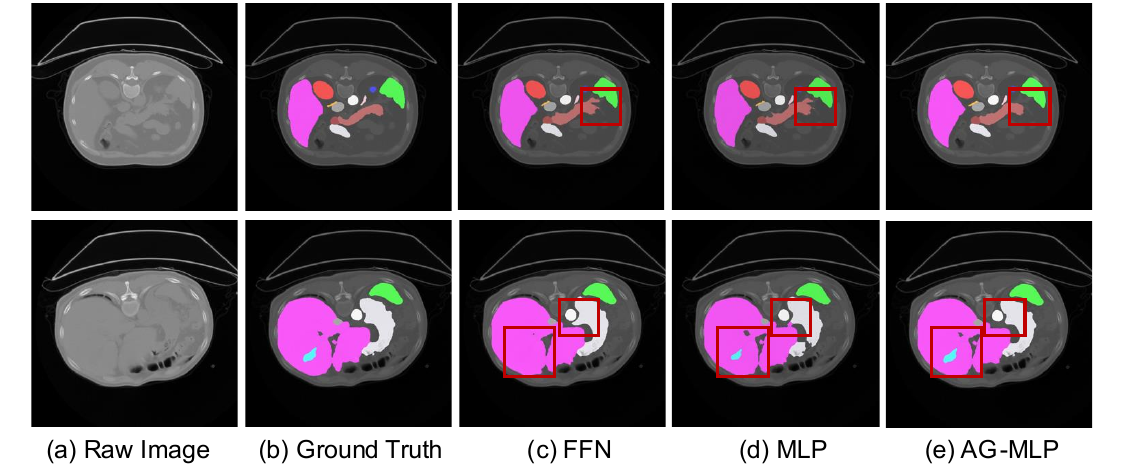}
\caption{Qualitative comparison among TransLK-NETR when they are incorporated with (c) FFN, (d) MLP, and (e) AG-MLP on the AMOS multi-organ segmentation task. Red boxes mark the regions where TransLK-Net incorporated with AG-MLP demonstrate better segmentation results than other networks.}
\label{vis4}
\end{figure}

To demonstrate the improvement of AG-MLP on segmentation performance over the FFN and standard MLP, we implemented an ablation study on AG-MLP in TransLK-Net. Employing AG-MLP in TransLK-Net improved segmentation performance with a slight increase in computational complexity compared with FFN and MLP (Table \ref{tab8} and Figure \ref{vis4}). Specifically, incorporating AG-MLP into TransLK-Net achieved $1.57$ and $1.2$ higher points in DSC with $5.2\%$ and $0.2\%$ more Params and $17.5\%$ and $0.6\%$ more FLOPs than incorporating FFN and MLP, respectively. Thus, the employment of AG-MLP improves the segmentation performance of these networks by enhancing their capabilities of learning spatial representations.

\begin{table}[!t]
\centering
\caption{Comparison of segmentation performance and computational complexity between AG-MLP and FFN, MLP in TransLK-Net on 2022 AMOS Multi-organ segmentation task. Performance was evaluated using DSC ($\%$; Mean$\pm$STD). The Params and FLOPs were evaluated using the input patches with the dimension of $96\times96\times96$.}
\label{tab8}
\resizebox{0.48\textwidth}{!}{
\begin{tabular}{c|c|cc}
\toprule
 Modules & DSC & Params (M) & FLOPs (G) \\
\midrule
FFN    & 88.22$\pm$13.66 & 38.94 & 73.31 \\
MLP    & 88.59$\pm$12.15 & 40.86 & 85.63 \\
AG-MLP & 89.79$\pm$11.46 & 40.95 & 86.17 \\
\bottomrule
\end{tabular}}
\end{table}

\subsubsection{Analysis of the Optimal Head Number}
PTLK and CTLK blocks employ a hyper-parameter, termed the head number $N$, in MHLK and DESA. It describes how channels are split and how convolutional layers and self-attention are applied. Specifically, splitting channels into multiple heads enables MHLK to apply different convolutional layers to avoid redundant feature extraction. MHLK will grow the kernel size $k$ with $N$ as $k=3+2*(N-1)$ to capture multi-scale features. Similarly, calculating self-attention scores in multiple heads ensures DESA focuses on different parts of input sequences, thus capturing richer features. To investigate its impact on segmentation performance and find the optimal value, we evaluated the performance of TransLK-Net when it employed different $N$. Specifically, we set the minimal number of $N$ as 2 to leverage the benefits of splitting channels to multiple heads. In contrast, splitting channels with a large head number leads to high computational complexity and difficulty in network optimization. Additionally, this $N$ should be set to make sure channels can be equally divided. Therefore, we investigated three potential numbers for $N$ ($N=2$, 3, and 4) (Table \ref{tab9}). 

Employing three heads ($N=3$) in TransLK-Net demonstrated the best segmentation performance, improving DSC scores by $2.28$ points than employing two heads ($N=2$), while slightly increasing $1.39\%$ Params and $4.7\%$ FLOPs, respectively. Employing convolution layers with more varying kernels enhanced models' capabilities of capturing more diverse features, thus improving segmentation performance. Additionally, employing convolutions with larger kernels slightly increased the computational complexity of these networks since these convolutions are applied within each head to avoid redundant feature extraction. Moreover, employing three heads in TransLK-Net achieved competitive performance with employing four heads ($N=4$) with fewer Params and FLOPs. Thus, we selected 3 as the optimal head number $N^*$ and split channels to three heads in MHLK and DESA.

\begin{table}[!t]
\centering
\caption{Comparison of segmentation performance and computational complexity among TransLK-Net employing different head numbers $N$ on 2022 AMOS Multi-organ segmentation task. Performance was evaluated using DSC ($\%$; Mean$\pm$STD). The Params (M) and FLOPs (G) were evaluated using the input patches with the dimension of $96\times96\times96$.}
\label{tab9}
\resizebox{0.48\textwidth}{!}{
\begin{tabular}{c|c|cc}
\toprule
Head $N$ & Mean DSC & Params (M) & FLOPs (G) \\
\midrule
2 & 87.51$\pm$12.96 & 40.43 & 82.32 \\
3 & 89.79$\pm$11.46 & 40.95 & 86.19 \\
4 & 89.54$\pm$11.86 & 41.78 & 92.27 \\
\bottomrule
\end{tabular}}
\end{table}

\section{Conclusion}
We propose a novel TransLK-Net for volumetric medical image segmentation by leveraging the benefits of large kernel convolutions and transformer self-attention to encode and decode features progressively and collaboratively. It employs the Progressively Entangled Transformer Large Kernel module and the Collaboratively Entangled Transformer Large Kernel module to capture mutually-calibrated local and global features. Subsequently, the Cross Entanglement Decoding block is built based on these two modules to enhance feature interactions and fusion. Additionally, AG-MLP is proposed to improve the capabilities of FFN and MLP in spatial feature learning. We evaluated TransLK-Net on four heterogeneous segmentation tasks, and it demonstrated superior segmentation performance with lower computation complexity compared to various SOTA methods. Due to their high generalizability and applicability, we believe that TransLK-Net has the potential to achieve promising segmentation performance on various medical image segmentation tasks.

\section*{Declaration of competing interests}
The authors declare that they have no known competing financial interests or personal relationships that could have appeared to influence the work reported in this paper.

\section*{Acknowledgments}
Computations were performed using the facilities of the Washington University Research Computing and Informatics Facility (RCIF). The RCIF has received funding from NIH S10 program grants: 1S10OD025200-01A1 and 1S10OD030477-01.

\section*{CRediT authorship contribution statement}
JY: conceptualization, methodology, formal analysis, writing the original draft, reviewing, and editing, visualization; DM: methodology, writing, reviewing, and editing, supervision; AS: methodology, writing, reviewing, and editing, supervision.

% Use the style of numbering in square brackets.
% If nothing is used, default style will be taken.
%\begin{enumerate}[a)]
%\item 
%\item 
%\item 
%\end{enumerate}  

%\begin{itemize}
%\item 
%\item 
%\item 
%\end{itemize}  

%\begin{description}
%\item[]
%\item[] 
%\item[] 
%\end{description}  

% Uncomment and use as the case may be
%\begin{theorem} 
%\end{theorem}

% Uncomment and use as the case may be
%\begin{lemma} 
%\end{lemma}



\bibliographystyle{cas-model2-names}

\bibliography{cas-refs}



\end{document}